\documentclass[final,5p,number,sort&compress]{elsarticle}
\usepackage{lineno}
\usepackage{float}
\usepackage{hyperref}
\usepackage{amsmath}
\usepackage{amssymb}

\usepackage{amsthm}
\usepackage{enumerate}
\usepackage{cleveref}

\usepackage{hyphenat}
\journal{Nuc.\ Instrum.\ Meth. A}

\begin{document}

\begin{frontmatter}

\title{A New Measurement of the $^6$Li(n,$\alpha$)t Cross Section at MeV Energies Using a $^{252}$Cf Fission Chamber and $^6$Li Scintillators}

\author[berk]{\corref{cor1}Leo E. Kirsch}
\ead{kirsch2@berkeley.edu}
\author[lanl]{M. Devlin}
\author[lanl]{S. M. Mosby}
\author[lanl]{J. A. Gomez}
\address[berk]{Nuclear Engineering Department,UC Berkeley, CA 94720, USA}
\address[lanl]{P-27, Los Alamos National Lab, Los Alamos, NM 87545, USA}

\cortext[cor1]{Corresponding Author}

\begin{abstract}
A new measurement is presented of the $^6$Li(n,$\alpha$)t cross section from 245 keV to 10 MeV using a $^{252}$Cf fission chamber with $^6$LiI(Eu) and Cs$_2$LiYCl$_6$:Ce (CLYC) scintillators which act as both target and detector.
Neutron energies are determined from the time of flight (TOF) method using the signals from spontaneous fission and reaction product recoil.
Simulations of neutron downscatter in the crystals and fission chamber bring $^6$Li(n,$\alpha$)t cross section values measured with the $^6$LiI(Eu) into agreement with previous experiments and evaluations, except for two resonances at 4.2 and 6.5 MeV introduced by ENDF/B-VII.1. 
Suspected neutron transport modeling issues cause the cross section values obtained with CLYC to be discrepant above 2 MeV.
\end{abstract}

\begin{keyword}
\texttt{cross section \sep 252Cf fission chamber \sep 6Li \sep scintillator \sep CLYC \sep LiI(Eu)}
\end{keyword}

\end{frontmatter}

\section{\label{sec:intro}Introduction}
Neutron detection arrays, conceptual fission reactor designs, and other applications rely on precise knowledge of the $^6$Li(n,$\alpha$)t reaction \cite{1748-0221-7-03-C03028,Serp2014308}.
However, new experiments and R-Matrix fits have led some nuclear data evaluators and experimental physicists to express doubts about previously reported $^6$Li(n,$\alpha$)t cross sections \cite{cite:HaleRMat, cite:DevlinDiffXS, PhysRevC.93.064612}.
Furthermore, it has been suggested that contradictions between observed abundances of $^6$Li and $^7$Li with Big Bang Nucleosynthesis modeling from cross section data could be explained by the existence of long-lived massive, negatively charged leptonic dark matter particles \cite{0004-637X-644-1-229,PhysRevD.76.121302}.

Previous measurements of the $^6$Li(n,$\alpha$)t cross section rely on $^6$LiI(Eu) \cite{J.NP_A.330.1.197910,J.PR.114.201.1959,J.PR.115.1707.59} and Li-glass detectors \cite{J.NIM_B.94.319.1994,J.NSE.71.205.7908}, while some use $^6$Li target foils \cite{cite:DevlinDiffXS,J.NSE.134.312.2000,NSR1956RI34}.
All previous procedures in the literature utilize charged particle beams to produce neutrons via $^1$H(t,n), $^2$H(d,n), $^3$H(p,n), $^7$Li(p,n), or spallation W(p,xn).
These measurements disagree as much as 20\% in the few MeV region.
The experiment introduced in this paper is the first $^6$Li(n,$\alpha$)t experiment known to the authors without a charged particle beam.
We test the performance of a $^{252}$Cf fission chamber neutron source with the conventional $^6$LiI(Eu) detector so that the measurements can contribute absolute $^6$Li(n,$\alpha$)t cross section data from 245 keV to 10 MeV for use in future evaluations.

This experiment also explores the quantitative reliability of Cs$_2$LiYCl$_6$:Ce scintillators (CLYC) by testing a $^6$Li enriched CLYC detector (C$^6$LYC) with the same fission chamber procedure to see if it can reproduce the standard $^6$Li(n,$\alpha$)t cross section.
Previous experiments have not been able to reproduce simulations of $^{35,37}$Cl(n,p) response in $^7$Li enriched CLYC detectors (C$^7$LYC) \cite{cite:D׳Olympia2014433}.
Poorly modeled neutron transport inside the crystal is likely the cause of the mismatch since there are not many measurements of $^{35,37}$Cl(n,*) from 100 keV to 14 MeV.
New results will give the CLYC development community some insight to proceed with optimization of crystal fabrication and experimental design. 

\section{\label{sec:expt}Experiment}
The procedure to precisely measure a cross section consists of three concepts: time of flight to determine neutron energy, identification of $^6$Li(n,$\alpha$)t events, and scattering simulations of the $^{252}$Cf neutron spectrum.

\begin{figure}
\centerline{\includegraphics[width=\linewidth]{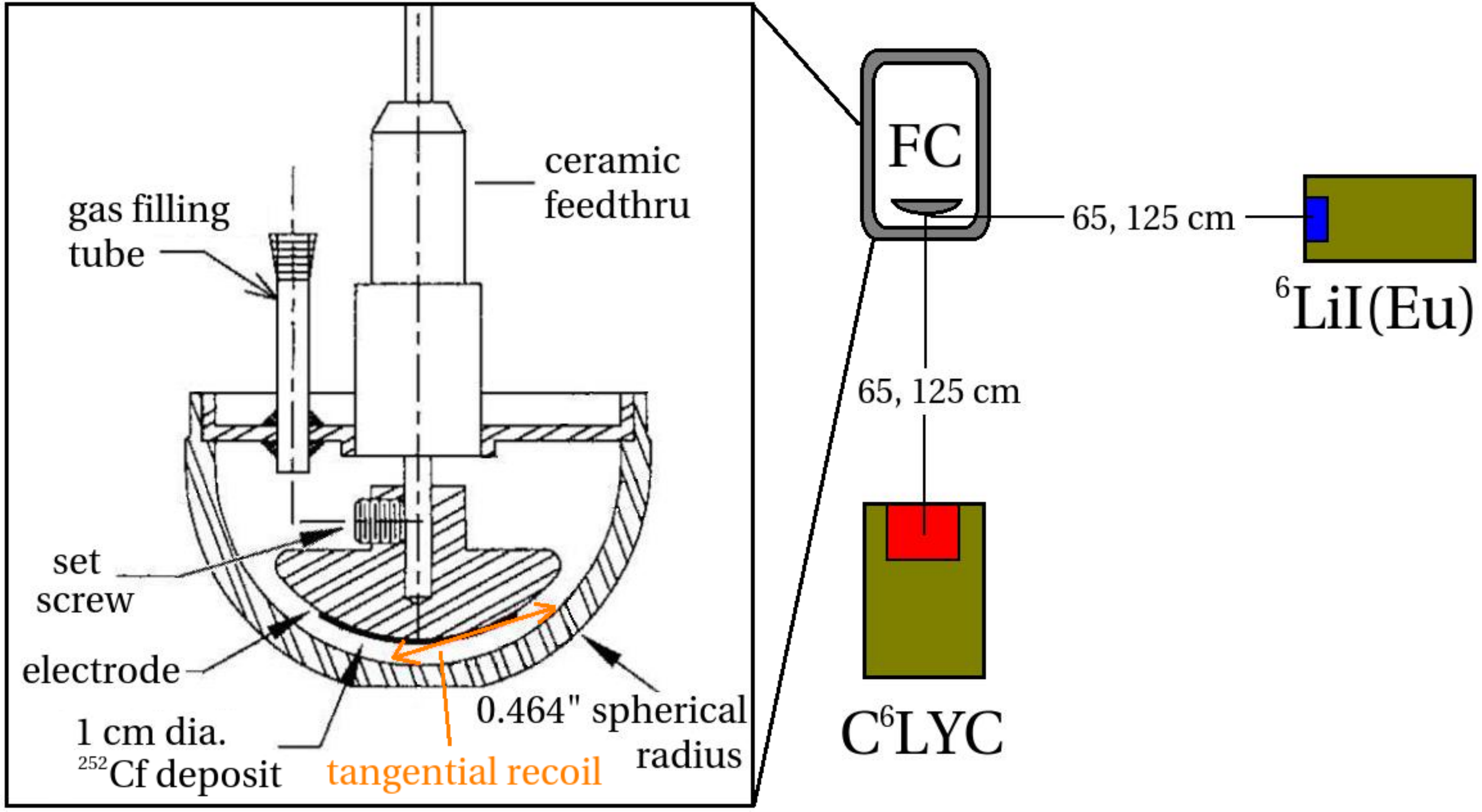}}
\caption{Experimental setup and schematic of the fission chamber. 
Back-to-back fission fragments cannot both recoil into the electrode due to its hemispherical shape.}
\label{fig:CADFC}
\end{figure}

The 1" diameter fission chamber\footnote{Double Contained Hemispherical Fission Chamber CF-53, model Q-6610-1, Rev.\ 2, manufactured by Oak Ridge National Laboratory.} was situated at the center of an otherwise empty 8$\times$6 meter room inside the Lujan Center of the Los Alamos Neutron Science Center for several months.
Roughly 0.345 $\mu$g of $^{252}$Cf initially resided on the rounded side of the enclosed hemispherical electrode. 
The $^{252}$Cf spontaneous fission fragments recoil through an electrically biased gas cell producing a signal. 
Fission fragments recoiling in a plane tangential to the $^{252}$Cf surface layer produce a larger signal since both fragments contribute to the Townsend avalanche as seen in Figure \ref{fig:CADFC}; hence, all fissions produce a signal.
The fragments emit an average of 3.76 prompt neutrons per fission for a total of approximately 795,000 neutrons per second.

Neutrons pass through 2-3 mm of the stainless steel capsule and into the open air room before potentially striking one of two lithium detectors.
In two iterations of this experiment, the detectors were 65 cm and 125 cm from the fission chamber.
All experimental and simulated data presented in this paper correspond to the 65 cm ``near configuration'' (except Figures \ref{fig:finalXS}, \ref{fig:closeXS}, and Table \ref{tab:xs}).
Due to the falloff in neutron flux with distance $(\sim 1/d^{2})$, various features are more prominent in the near configuration. 
Analysis of the 125 cm ``far configuration'' data uses the same procedure, but for brevity the plots are omitted and are available upon request.
Data were collected for 27 days in each configuration.
The $^6$LiI(Eu) and C$^6$LYC scintillators performed duties of both target and detector: $^6$Li(n,$\alpha$)t charged particle reaction products recoil inside the crystals generating a luminescent response.
Photomultiplier tubes (PMTs) attached to the crystal faces convert the light into electrical signals.

Signal output from the fission chamber and detectors transmit directly into separate channels of a 14-bit, 500 MS/s digitizer\footnote{Model VX1730. Manufactured by CAEN S.p.A., Viareggio, LU 55049, Italy.}.
Each input independently self-triggers using leading edge discrimination.
Waveform pre-processing removes the $^{252}$Cf alpha decay and fission fragment beta decay; while waveform post-processing uses a constant fraction discrimination (CFD) algorithm to assign pulse\hyp{}height\hyp{}independent timestamps.
The digitizer records signal output every 2 ns, but interpolation of CFD output yields sub-nanosecond zero crossings.
Timing resolution is a function of detector pulse height and can reach standard deviations as low as 1.5 ns for $^6$LiI(Eu) and 0.5 ns for C$^6$LYC in this experiment.
A data analysis code organizes coincident fission chamber and detector output into an event structure containing timestamp and pulse height information.
An event satisfies the coincidence condition if either detector triggered as early as 75 ns before or as late as 500 ns after fission.
Section \ref{sec:absNorm} addresses dead time issues resulting from the high fission rate.

The time of flight (TOF) of a neutron is the time difference between the emission signal from the fission chamber and the reaction signal from the scintillators' PMTs.
Varying cable lengths, digital signal post-processing, and other experimental factors obfuscate an absolute TOF measurement. 
Therefore, a timing calibration ensures that prompt photons from fission arrive precisely at time $t = d / c$, where $d$ is the distance from the $^{252}$Cf fission deposit to the center of the detector active volume.
No attempt was made to absolutely calibrate the time offset.
The relativistic conversion to neutron kinetic energy $E_n$ is
\begin{equation}
\label{eq:En}
E_n = \left( \frac{1} { \sqrt{1 - \left( \frac{d} {c \cdot \textrm{TOF}} \right)^2 } } - 1 \right) m c^2,
\end{equation}
where $m$ is the rest mass of the neutron.
The TOF uncertainty is the aforementioned detector pulse height dependent timing resolution.
The distance uncertainty is a combination of digital laser sight systematic uncertainty, detector half thicknesses, machining tolerances, and neutron source spatial distribution: 5 mm for $^6$LiI(Eu) and 8.5 mm for C$^6$LYC.

Pulse height magnitude typically reflects the sum of a reaction Q-value and incident neutron kinetic energy.
However, different recoiling charged particles generate different amounts of light in the stopping process.
For example, prompt fission $\gamma$-rays Compton scatter to produce electrons of a few hundred keV, the $^6$Li(n,$\alpha$)t reaction produces tritons and alphas up to 13 MeV, $^{35}$Cl(n,p) produces protons up to 10 MeV, and $\gamma$-rays from $^{127}$I(n,inl) produce electrons of a couple hundred keV in photoelectric absorption.

\begin{figure}
\centerline{\includegraphics[width=\linewidth]{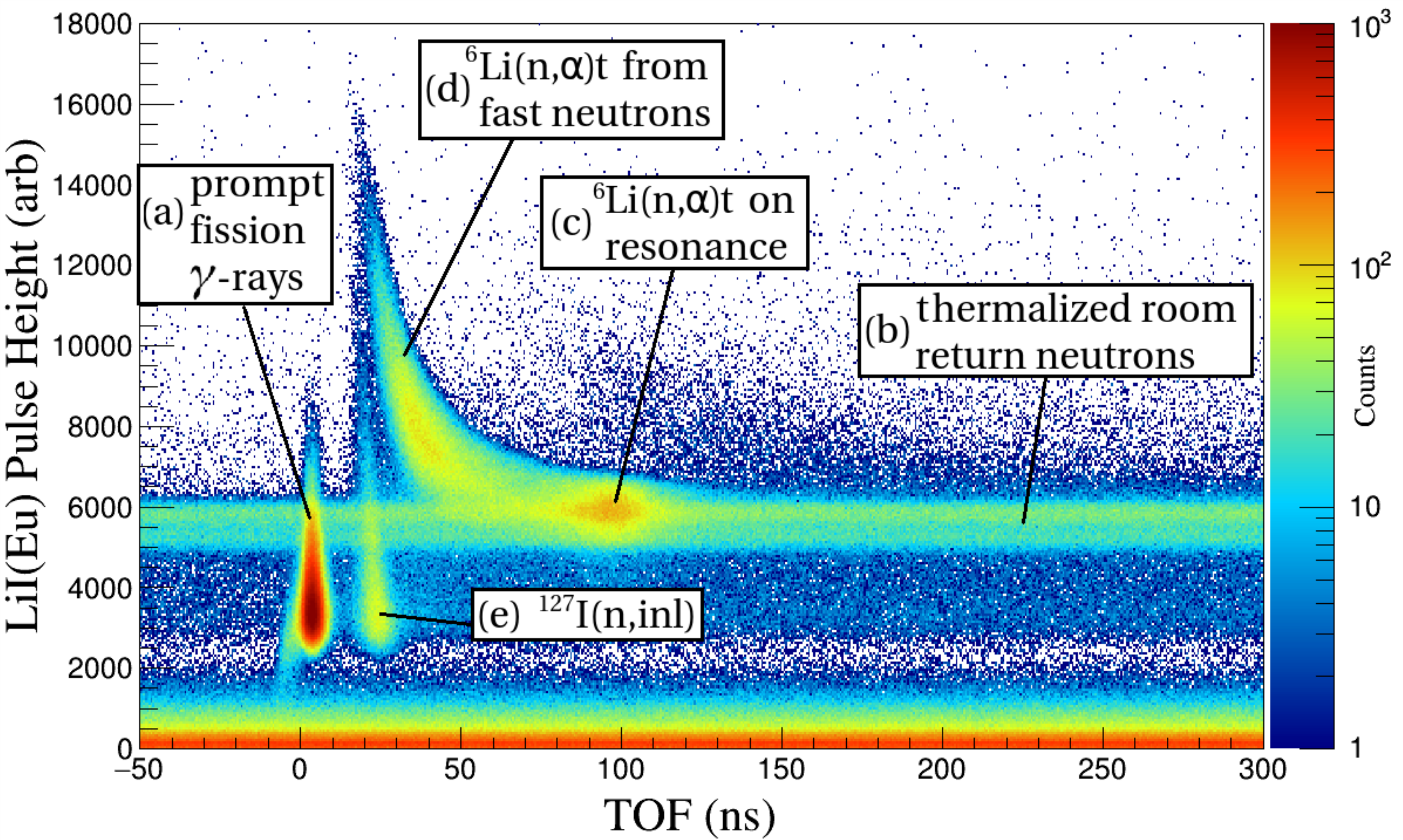}}
\caption{$^6$LiI(Eu) experimental TOF spectrum. 
Additional descriptions corresponding to the letters given in the text.
The experimental and simulated plots in this paper correspond to the 65 cm near configuration except where otherwise noted.
}
\label{fig:nTOFPHLi}
\end{figure}
The $^6$LiI(Eu) detector\footnote{Model 25.4B/1.5LiI(neg). Manufactured by ScintiTech Inc., Shirley, MA 01466, USA. PMT Photonis XP2042.} has an active crystal\footnote{Manufactured by Cryos-Beta Ltd., 60 Lenina Avenue, Kharkov 61001, Ukraine.} volume measuring 25 mm in diameter by 3 mm in thickness. 
Production of $^6$LiI crystals began in the early 1950's \cite{cite:PhysRev.82.749,cite:PhysRev.85.919,cite:firstEu}.
After several decades of testing, many dependable descriptions of fast neutron pulse height spectra exist \cite{cite:Ophel195845,cite:Johnson196961}.
Figure \ref{fig:nTOFPHLi} shows the pulse height vs.\ TOF spectra for $^6$LiI(Eu). 
In this experiment there are many noteworthy features:
\begin{enumerate}[(a)]
\item Events within the vertical ``teardrop'' shape calibrated to time $t = d / c$ are prompt fission $\gamma$-rays Compton-scattering off electrons in the crystal.
\item Events within the horizontal band at pulse height 5500 stretching over all times are $^6$Li(n,$\alpha$)t reactions from thermalized room return neutrons reaching the crystal. 
This reaction has a Q-value of +4.78 MeV.
\item The surge of events at 95 ns with pulse height centered slightly above the thermal band are $^6$Li(n,$\alpha$)t reactions near the 240 keV resonance from unscattered fission neutrons. 
\item Events within the ``banana'' shape that extends from the resonance toward 12 ns with increasing pulse height are also $^6$Li(n,$\alpha$)t reactions from unscattered fission neutrons.
This band of events exhibits a kinematic curve because high energy neutrons that arrive earlier produce a larger pulse height.
\item Events within the ``curved teardrop'' around 25 ns are photoabsorption or Compton scattering of the $\gamma$-rays from $^{127}$I(n,inl). 
The pulse height separation of this band with $^6$Li(n,$\alpha$)t arises from the different reaction Q-values.
\end{enumerate}
Events from $^{127}$I(n,inl) begin to appear below approximately 33 ns corresponding to incident neutron energies above 2.0 MeV for the near configuration via Equation \eqref{eq:En}. 
This energy onset is much higher than the 1st excited state of $^{127}$I at 57.6 keV.
Events depositing less than 2.0 MeV do not trigger the pulse height threshold of the digitizer data collection software except by random coincidence.


The C$^6$LYC scintillator\footnote{Model CLYC-627-1A.u PMT Hammatsu R6231-100.} has an active volume measuring 25.4 mm in diameter by 10 mm in thickness.
Production of CLYC began in 1999 \cite{cite:Combes1999299}, and only recently have other groups explored the crystal's response to fast neutrons \cite{D'Olympia2012140}.
\begin{figure}
\centerline{\includegraphics[width=\linewidth]{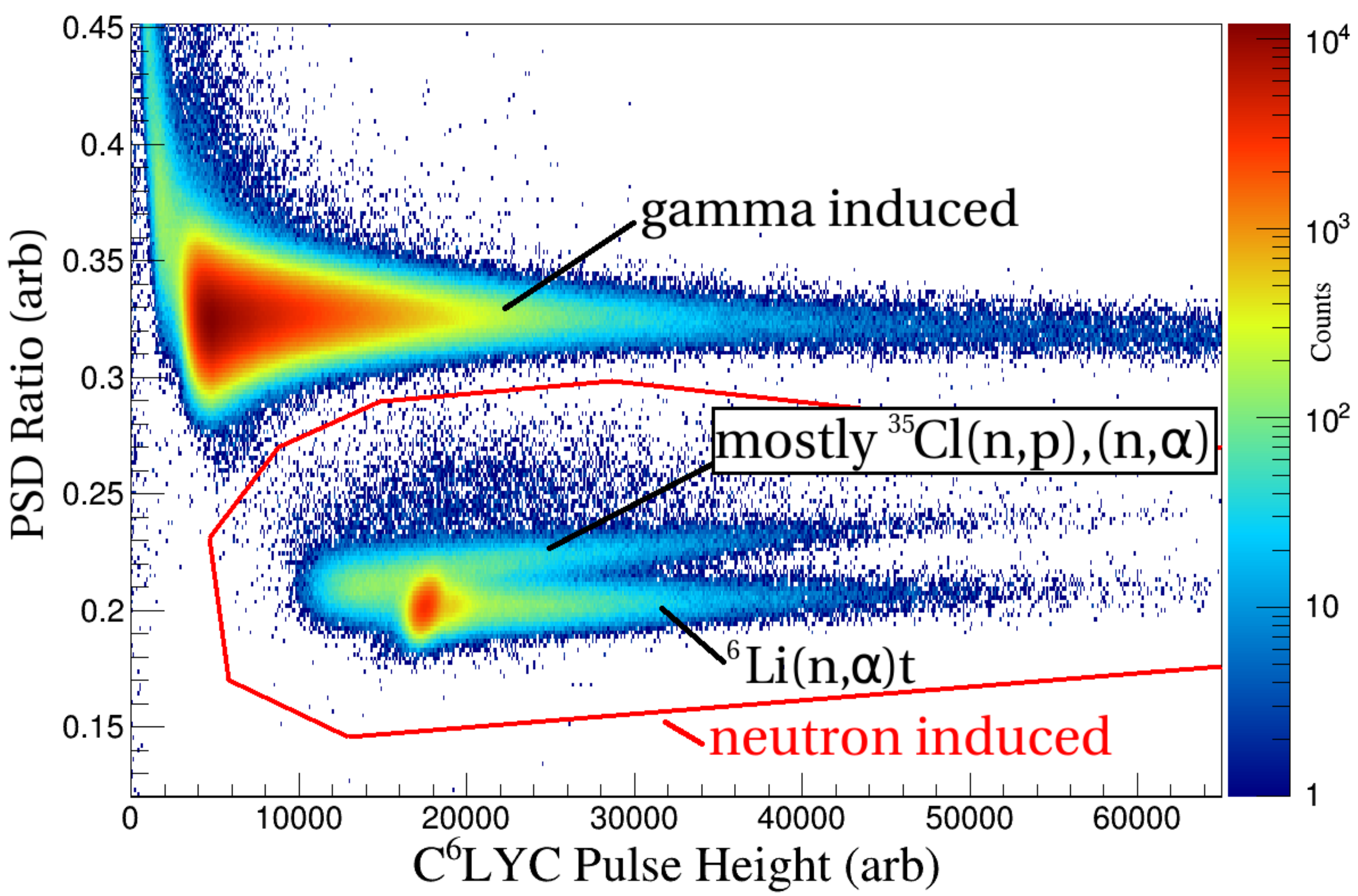}}
\caption{C$^6$LYC pulse shape discrimination for the full spectra of neutrons and $\gamma$-rays from $^{252}$Cf spontaneous fission.
The three neutron induced bands that arise from $^6$Li(n,$\alpha$)t, $^{35}$Cl(n,p), and $^{35}$Cl(n,$\alpha$) are not separable from each other with PSD.
}
\label{fig:PSD}
\end{figure}
The C$^6$LYC detector has the benefits of pulse shape discrimination (PSD), shown in Figure \ref{fig:PSD}.
The PSD process uses the scintillation decay profile to identify if the event is a $\gamma$-ray induced electron recoil or a neutron induced proton or alpha recoil.
Electron recoil has a shorter scintillation decay time and therefore has a larger PSD ratio
\begin{equation}
PSD  = \frac{h_{short}}{h_{long}},
\end{equation}
where $h_{short}$ and $h_{long}$ are integrated pulse heights of C$^6$LYC PMT output over short and long intervals, respectively. 
Pulse height integration begins 112 ns before the pulse height trigger, and short and long time intervals have durations of 120 and 600 ns, respectively.
There was no optimization attempt of PSD short and long time intervals. 
As Figure \ref{fig:PSD} shows, neutron induced events of all integrated pulse heights are almost completely disconnected from $\gamma$-ray induced events using PSD. 
Additionally, TOF information separates neutrons and $\gamma$-rays since the majority of $\gamma$-rays are prompt products of fission and arrive at the detector much earlier than fission neutrons.
Hence, there was minimal ambiguity as to which events were neutrons or $\gamma$-rays for this experiment's neutron energy range and source to detector distances.

\begin{figure}
\centerline{\includegraphics[width=\linewidth]{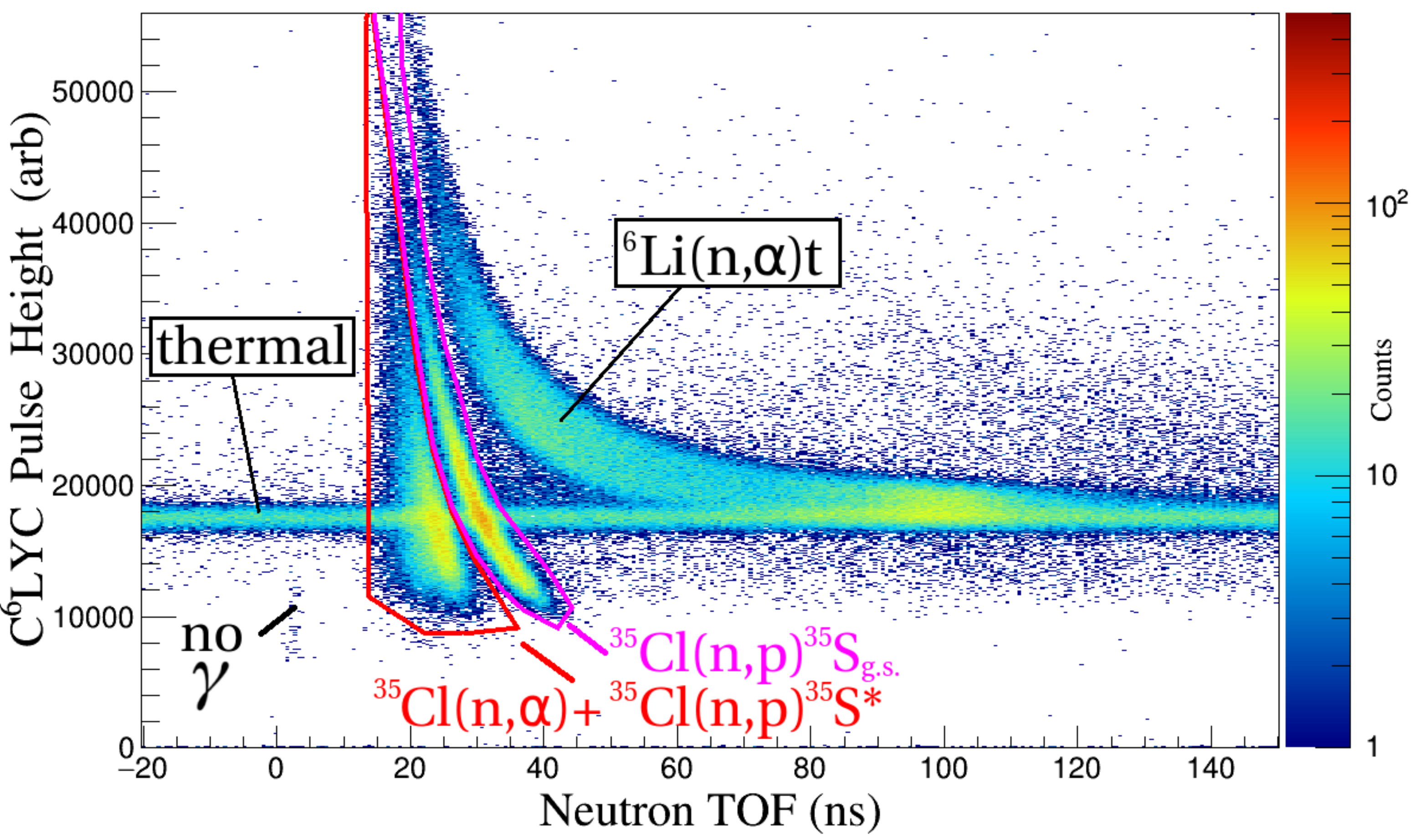}}
\caption{C$^6$LYC experimental TOF spectrum.
The $^{35}$Cl(n,p)$^{35}$S reaction has Q-value +615 keV but does not have a large cross section until the neutron has over 1 MeV. 
Neutron pulse shape discrimination of Figure \ref{fig:PSD} ensures $\gamma$-ray events do not appear.
}
\label{fig:nTOFPHCLYC}
\end{figure}

Figure \ref{fig:nTOFPHCLYC} shows the pulse height vs.\ TOF spectrum for C$^6$LYC. 
This spectrum has features similar to that of $^6$LiI(Eu) with the addition of two new kinematic curves.
Events within the middle ``banana'' are $^{35}$Cl(n,p)$^{35}$S$_{\textrm{g.s.}}$ reactions where $^{35}$S$_{\textrm{g.s.}}$ is the ground state of $^{35}$S.
Events within the bottom ``banana'' are the combination of $^{35}$Cl(n,$\alpha$)$^{32}$P and $^{35}$Cl(n,p)$^{35}$S* reactions where $^{35}$S* are excited states starting at 1.57 MeV.
This band may also contain events from other reactions with charged particle products such as $^{37}$Cl(n,p),(n,$\alpha$), or $^{35}$Cl(n,d).

\begin{figure}
\centerline{\includegraphics[width=\linewidth]{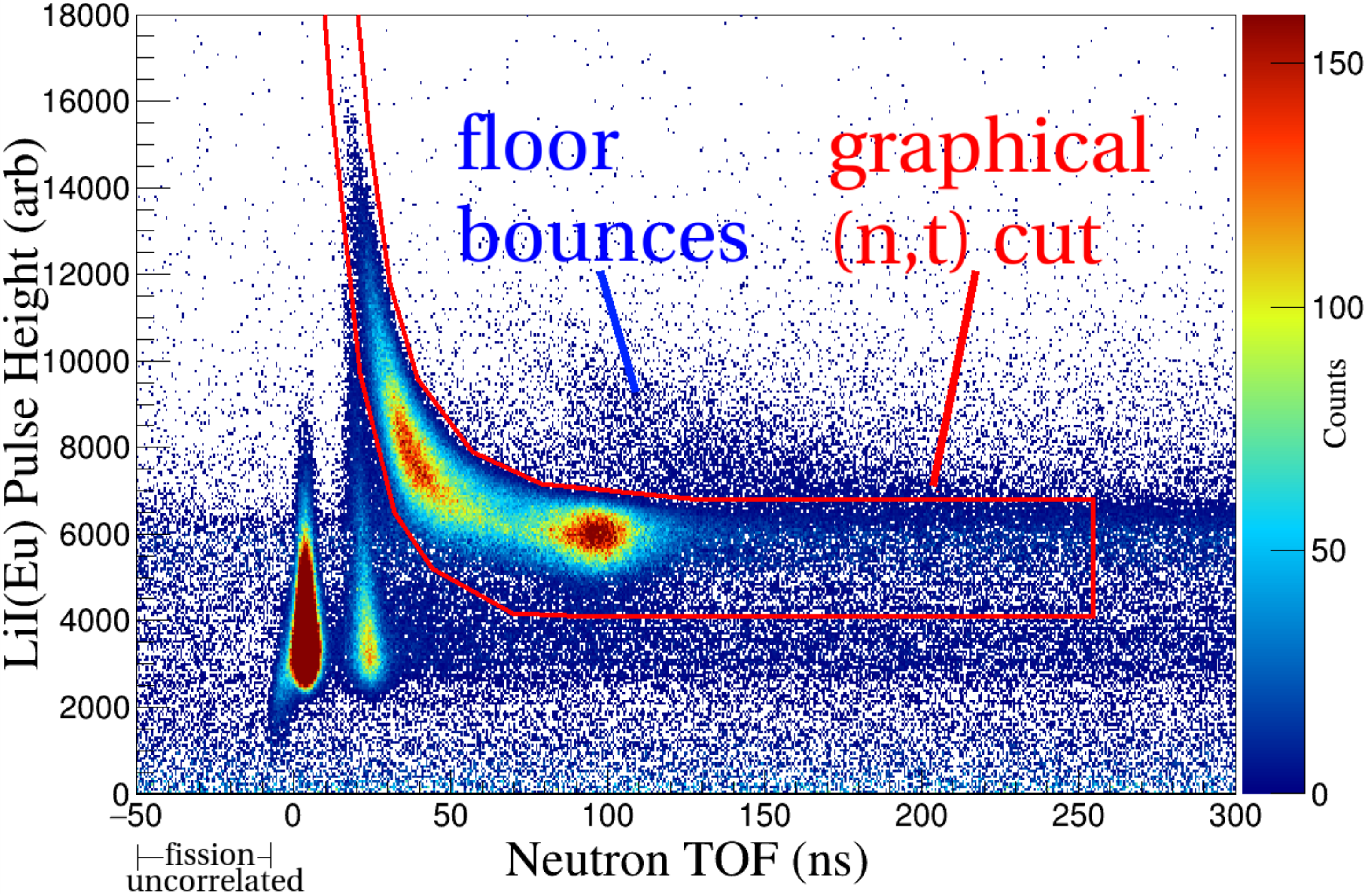}}
\centerline{\includegraphics[width=\linewidth]{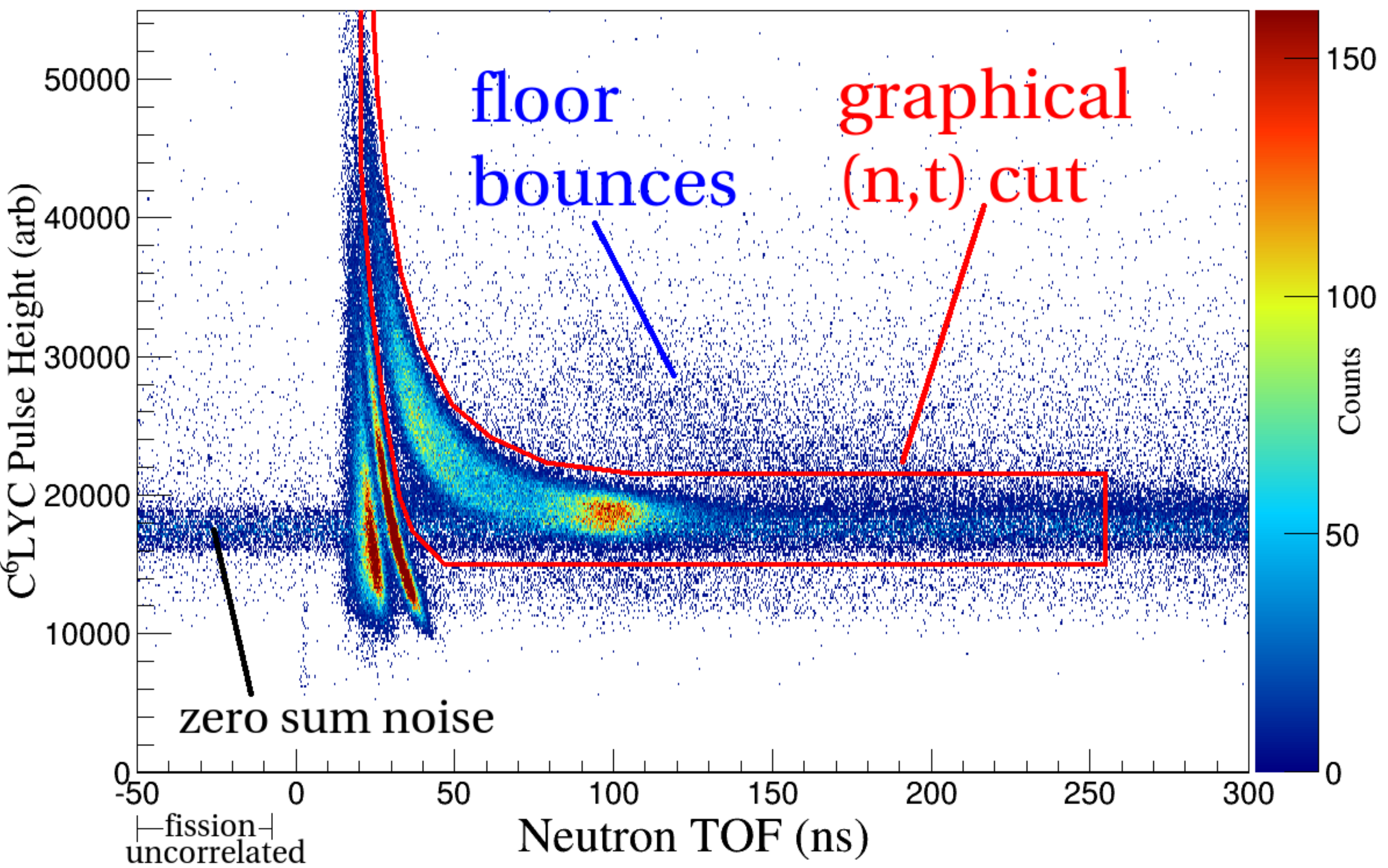}}
\caption{Background subtracted pulse height vs.\ neutron TOF: top $^6$LiI(Eu), bottom C$^6$LYC. 
The fission uncorrelated region is the background subtraction source.
The graphical cut selects $^6$Li(n,$\alpha$)t events for cross section calculations. 
Faint room return neutron bands show that neutrons taking a single bounce off the floor and deflecting into the detectors retain the majority of their energy but arrive noticeably later in time.
}
\label{fig:bksub}
\end{figure}

Obtaining the net number of $^6$Li(n,$\alpha$)t events, $n_n$, requires a good understanding of fission-induced background.
Neutrons that thermalize in surrounding material and migrate to the detectors contribute a constant background.
Neutron and $\gamma$-ray events with recorded ``negative TOF'' spawn from earlier fissions not associated with the most recent trigger.
Figure \ref{fig:bksub} shows pulse height vs.\ TOF spectra after subtraction of background events averaged from -49 ns to -9 ns.
Subtracting background this way removes the fully thermalized neutron bands, but the structure of room return neutrons persists.
The pre-fission background subtraction does not remove single scatter neutrons that maintain a timing correlation with the fission trigger.

Section \ref{sec:Corr} investigates the geometric quality of the graphical cut in Figure \ref{fig:bksub}. 
Specifically, Sections \ref{ssec:Room}, \ref{ssec:FC} and \ref{ssec:Xtal} address the fact that the graphical cut includes some intrusive $^6$Li(n,$\alpha$)t events in which neutrons first scatter on materials inside and outside the detectors.
Moreover, Section \ref{ssec:resp} addresses the fact that the graphical cut excludes some legitimate $^6$Li(n,$\alpha$)t events due to particle ejection from the crystal and the broadening of the kinematic curve due to detector pulse height and timing resolutions.


\section{\label{sec:absNorm}Absolute Normalization}

The expression for the absolute $^6$Li(n,$\alpha$)t cross section $\sigma$ is
\begin{equation}
\label{eq:xs}
\sigma(E_n) = \frac{n_n(E_n)}{N_{f,tot} N_{^6\textrm{Li}} \phi_{det} n_f(E_n) },
\end{equation}
where $n_n$ is the net number of background subtracted $^6$Li(n,$\alpha$)t events detected for incident neutron energy $E_n$, 
$N_{f,tot}$ is the total number of neutrons emitted during the count time of the experiment, 
$N_{^6\textrm{Li}}$ is the number of $^6$Li nuclei in each crystal,
$\phi_{det}$ is the path length estimate of particle flux for the geometries of this experiment, 
and $n_f$ is the relative number of neutrons emitted from fission at energy $E_n$.

The detectors and fission chamber do not register events for the entirety of the experiment. 
The quantity $N_{f,tot}$ includes dead time and down time losses which effectively reduce the number of usable neutrons:
\begin{equation}
N_{f,tot} = f_{live} f_{rec} \bar{\nu}_p \int_{t_i}^{t_f} dt' A_{252,i} 2^{-t'/t_{1/2}},
\label{eq:fissInt}
\end{equation}
where 
$f_{live}$ is the fraction of time the digitizer boards are live, 
$f_{rec}$ is the fraction of time the data acquisition system is recording,
$\bar{\nu}_p = 3.7590 \pm 0.0047$ is the average number of prompt neutrons emitted from $^{252}$Cf spontaneous fission, 
$t_i$ and $t_f$ are the start and end times of this experiment respectively, 
$A_{252,i}$ is the initial $^{252}$Cf fission rate, 
and 
$t_{1/2} = 2.645 \pm 0.008 $ yr is the $^{252}$Cf half-life.
The $^{252}$Cf fission rate changes slightly over the course of the experiment making the integration in Equation \eqref{eq:fissInt} necessary.
The run time ($t_f - t_i$) for the setup where the detectors were in the far configuration (125 cm) was $26.765 \pm 0.044$ days, causing approximately a 1.9\% decrease in the $^{252}$Cf fission rate.

\begin{figure}
\centerline{\includegraphics[width=\linewidth]{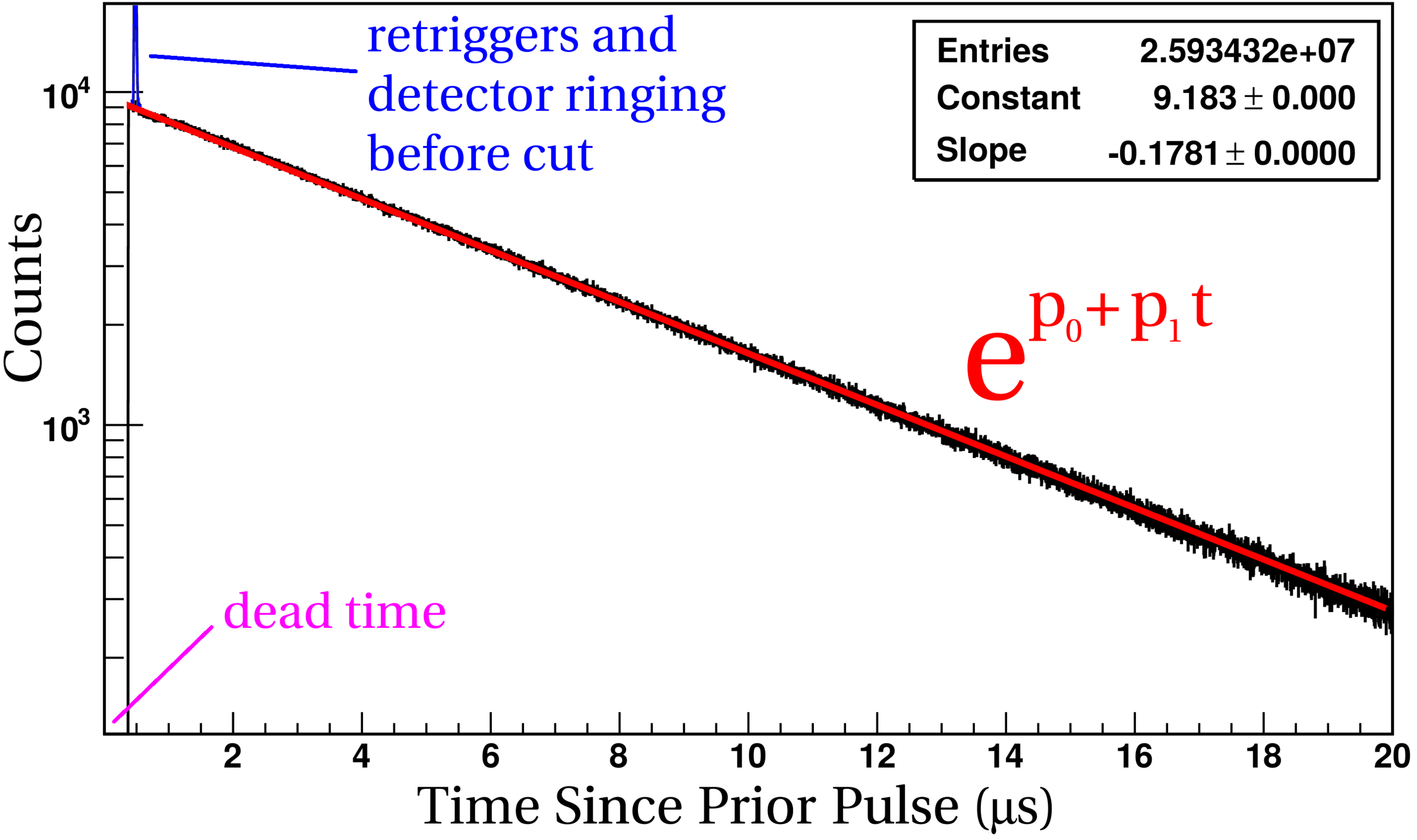}}
\caption{Interevent time of fission. 
Data taken on 3/16/17 with measured total fission rate $p_1=$ 178,100 f/s. 
The number of counts measured for a given interevent time is proportional to the probability of observing zero events during that time using a Poisson process.
}
\label{fig:fissRate}
\end{figure}

Figure \ref{fig:fissRate} shows the results of the fission chamber calibration and reveals that the digitizer board has some inherent dead time as there are seemingly no events until at least 372 ns after each pulse.
After subtracting out the effects of tail retriggering and detector ringing, the total number of entries in this calibration is $2.5932 \times 10^7$.
The integral of the exponential fit of counts vs.\ interevent time ($e^{p_0 + p_1 t}$ in Figure \ref{fig:fissRate}) is $2.7317 \times 10^7$.
This integral would equal the total number of entries if digitizer board could transfer data without breaks, but the high $^{252}$Cf pulse rate results in many unregistered events during this time.
However, Equation \eqref{eq:fissInt} uses the ratio of the number of calibration entries to integrated fit, referred to as live time fraction $f_{live} = 0.949 \pm 0.005$, which appropriately manages the high pulse rate.

The data acquisition system quickly generates large run files due to the high fission rate. 
An automated code stops the collection after 10 minutes of recording data to save the current file, begin a new run, and reinitialize parameters.
This down time is 1-2 seconds, and therefore the fractional recording time is $f_{rec} = 0.9975 \pm 0.0008$.

In practice, the fission chamber measures $^{252}$Cf fission as well as fission of various impurities and daughter nuclei:
\begin{equation}
A_{tot,i} = A_{252,i} / f_{252},
\end{equation}
where $A_{i,tot}$ is the initial total fission rate from all isotopes and $f_{252}$ is the fraction of fissions from $^{252}$Cf.
Table \ref{tab:cf} provides a summary of the mass assay of curium and californium isotopes from the fission chamber manufacturer and predicts a purity of $f_{252} = 0.9972 \pm 0.0005$ on the start day of the experiment.
\begin{table}
\centering
\caption{Projected Fission Rates of Californium Isotopes}
\resizebox{\linewidth}{!}{%
\begin{tabular}{c c c c c}
  \hline\hline
  Isotope & 12/10 Assay ($\mu$g) & $t_{1/2}$ (yr) & SF (\%) & 7/16 Fiss.\ Rate(f/s) \\
  \hline
  $^{246}$Cm & 0.006    & 4706   & 0.03   &  0.3     \\
  $^{248}$Cm & 0.206    & 3.48e5 & 8.39   &  18.1    \\
  $^{249}$Cf & 0.201    & 351    & 5.0e-7 &  0.053   \\
  $^{250}$Cf & 0.264    & 13.08  & 0.077  &  610     \\
  $^{252}$Cf & 1.608    & 2.645  & 3.09   &  2.264e5 \\
  $^{254}$Cf & $<$0.001 & 0.166  & 99.69  &          \\    
  \hline\hline
\end{tabular}}
\label{tab:cf}
\end{table}
The manufacturer also provides a separate analysis of the total fission rate and states that their mass assay values are 5-10\% too high.
This rate discrepancy does not greatly affect the purity levels.
The authors of this paper performed a separate calibration of the total fission rate in March 2017, over six years after the manufacturer's original December 2010 assay. 
Figure \ref{fig:fissRate} shows results from the new calibration and projects a total fission rate of $211,400 \pm 130$ f/s on the July 2016 start date of the experiment, confirming the projection of manufacturer's mass assay as 7\% too high.

The number of $^6$Li nuclei $N_{^6\textrm{Li}}$ in each crystal is 
\begin{equation}
N_{^6\textrm{Li}} = \pi R^2 T \rho \cdot N_{A} f_{^6\textrm{Li}} / M_X,
\end{equation}
where $R$, $T$, $\rho$, and $M_X$ are crystal radius, thickness, density, and molar mass, respectively, 
$N_A$ is Avagadro's number, 
and  
$f_{^6\textrm{Li}}$ is the $^6$Li enrichment.
The uncertainties for these quantities are quite low except for radius and thickness which have respective tolerances of 0.05 and 0.1 mm as quoted from the manufacturer.

A detector that collects particles from a distributed source does not have a well-defined solid angle.
Also, particles that graze the corners of the crystal do not have the same interaction probability as particles passing through the entire thickness.
Simulations using the Monte Carlo N-Particle transport code MCNP$^{\textregistered}$ \cite{cite:MCNP} determined the path length estimate of particle flux $\phi_{det}$ for the geometries of this experiment. 
Again, the most sensitive quantities are crystal radius and thickness of the crystal.
For the setup where $^6$LiI(Eu) was in the far configuration, $\phi_{det} = 5.107 \pm 0.005 \times 10^{-6} $ n/cm$^2$ per source neutron, meaning that on average, only five out of one million neutrons pass through the entire thickness of the crystal.
However, neutrons scatter in the fission chamber and the crystals themselves. 
Sections \ref{ssec:FC} and \ref{ssec:Xtal} address these scattering corrections.

The most up to date $^{252}$Cf prompt fission neutron spectrum $n_f(E_n)$ is the Mannhart evaluation \cite{cite:cfSpec} retrievable from the Evaluated Nuclear Data File (ENDF) database \cite{Chadwick20112887}.
The evaluation provides relative standard deviations varying from 1-3\% in the region from 100 keV to 10 MeV from a global fit to available data with a simple model of the shape of the neutron spectrum.
These uncertainties are much lower than the evaluated $^6$Li(n,$\alpha$)t cross section uncertainties, and do not contribute to the uncertainty in absolute normalization since $n_f$ is normalized to unity.

In total, the absolute normalization has an uncertainty of 3.6\% for $^6$LiI(Eu) in the far configuration.
The largest contributors to the uncertainty are crystal radius and thickness.
Both the number of $^6$Li nuclei and subtended neutron flux scale quadratically with radius.
For $^6$LiI(Eu), the uncertainty in crystal thickness dominates because the tolerance is a larger fraction of the whole dimension. 
The correction factors of the following section have comparable systematic uncertainties.

\section{\label{sec:Corr}Correction Factors }
Neutrons emitted from a fission chamber react with the environment differently than the collimated neutrons of previous beam experiments.
This section presents the measurement constraint due to room return, two correction factors for neutron downscatter in the fission chamber and crystal, and one correction factor for reaction product escape from the walls of the thin crystals. 


\subsection{\label{ssec:Room}Effect of Room Return}
More neutrons pass through the detectors than expected from a na\"ive conversion of the Mannhart $^{252}$Cf energy spectrum to TOF spectrum. 
Neutrons not initially moving in the direction of the detectors bounce off the room and react at times later than they would on a direct path from fission chamber to detector.
For example, the shortest path from fission chamber to the concrete floor (93 cm below) to detector is 225 cm.
Neutrons with kinetic energy of a few MeV on this path induce events which enter the graphical cut of Figure \ref{fig:bksub}.
However, the room walls are too far for neutron reflections to induce events which enter the graphical cut.

\begin{figure}
\centerline{\includegraphics[width=\linewidth]{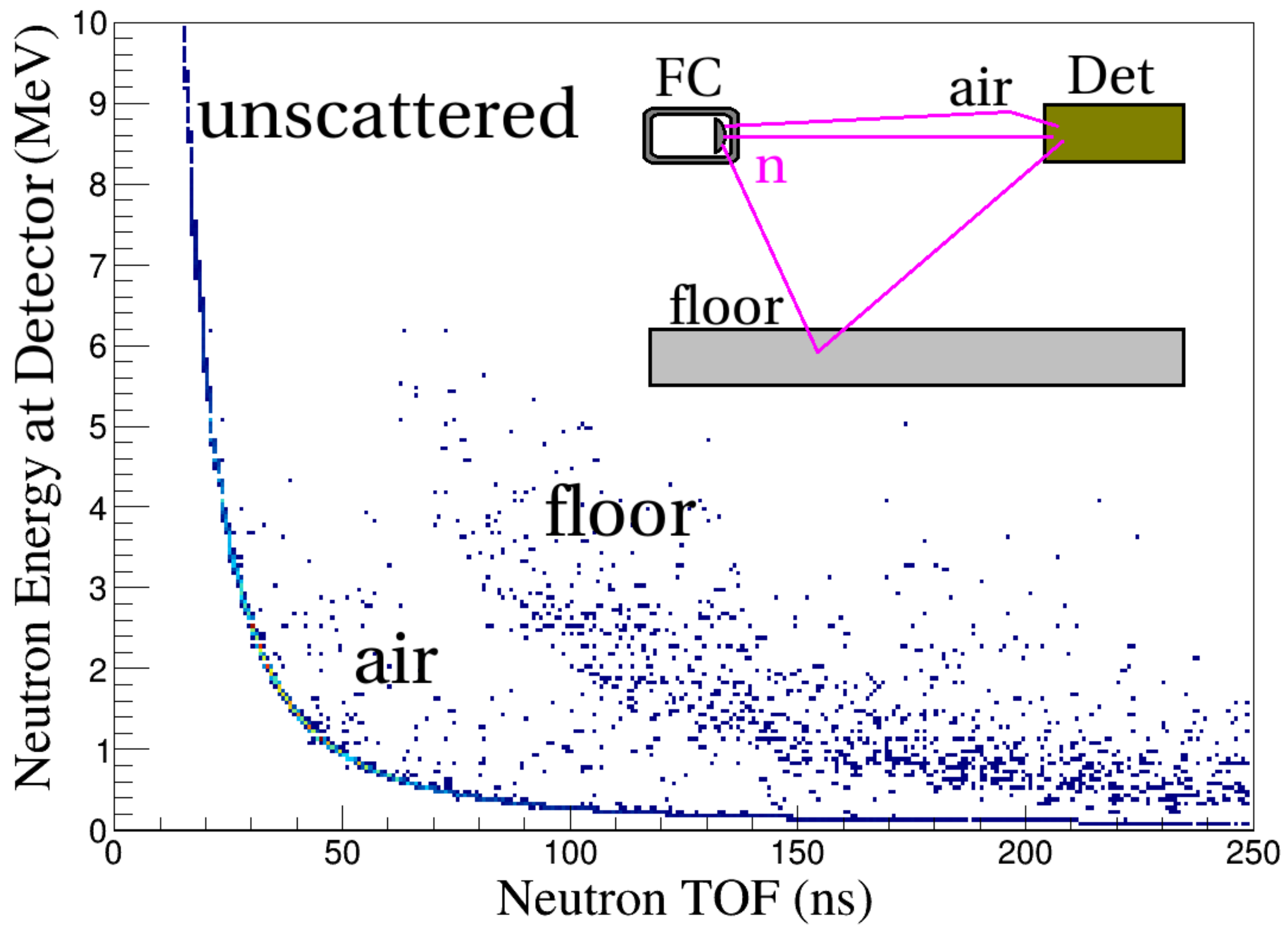}}
\caption{Simulation of room return including concrete floor and surrounding air.
The floor bounce band disappears when void replaces the concrete floor material.
The air bounce band disappears when void replaces the air material.}
\label{fig:nBounce}
\end{figure}

Figure \ref{fig:nBounce} shows results from an MCNP\textsuperscript{\textregistered} simulation which contains the distributed neutron source, the concrete floor, surrounding air, and a flux tally at the location of the detector. 
This simulation omits fission chamber and detector geometry.
The same floor bounce band appears as first noted in the experimental spectra of Figure \ref{fig:bksub}.
Overlap of the floor bounce and unscattered bands occurs when the energy difference is less than twice the experimental pulse height resolution.
For instance, $^6$LiI(Eu) has pulse height resolution approximately 1 MeV so the simulation predicts overlap for TOF $>$ 95 ns which matches the experimental values.
Then, according to Equation \eqref{eq:En}, the concrete floor does not interfere for $E_n > 245$ keV.

Figure \ref{fig:nBounce} also reveals an air bounce band which enters the graphical cut for all TOF. 
However, air bounces cause less than a 0.5\% change to flux at any energy.

\subsection{\label{ssec:FC}Effect of Fission Chamber Scatter}
\begin{figure}
\centerline{\includegraphics[width=0.75\linewidth]{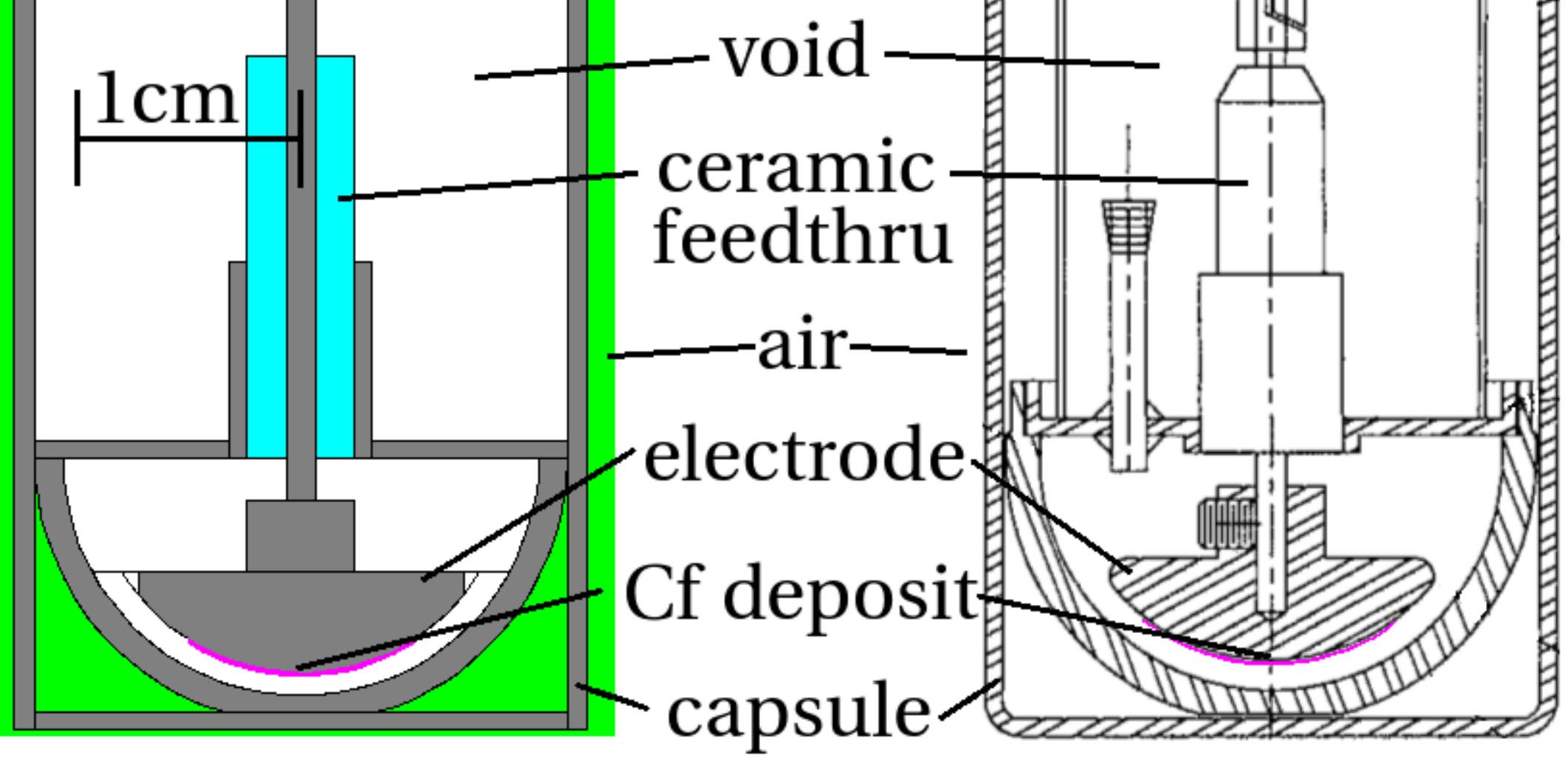}}
\caption{Simulation geometry of the fission chamber and $^{252}$Cf deposit. 
Trivial geometric features omitted from simulation for the sake of computation time. 
}
\label{fig:vertFC}
\end{figure}
The $^{252}$Cf fission chamber is not physically large enough to significantly alter neutron TOF.
Scattering converts fast neutron source particles into slow neutron source particles.
Figure \ref{fig:vertFC} shows that the dimensions of the fission chamber are of the order 1 cm.
Neutrons with kinetic energy 240 keV experience a 1.5 ns maximum deviation from the path length increase of a single scatter.
This change to TOF is greatest at low energy, but is always within experimental timing resolution.
Therefore, the fission chamber geometry modifies the neutron flux magnitude but not the relationship between energy and TOF.

\begin{figure}
\centerline{\includegraphics[width=\linewidth]{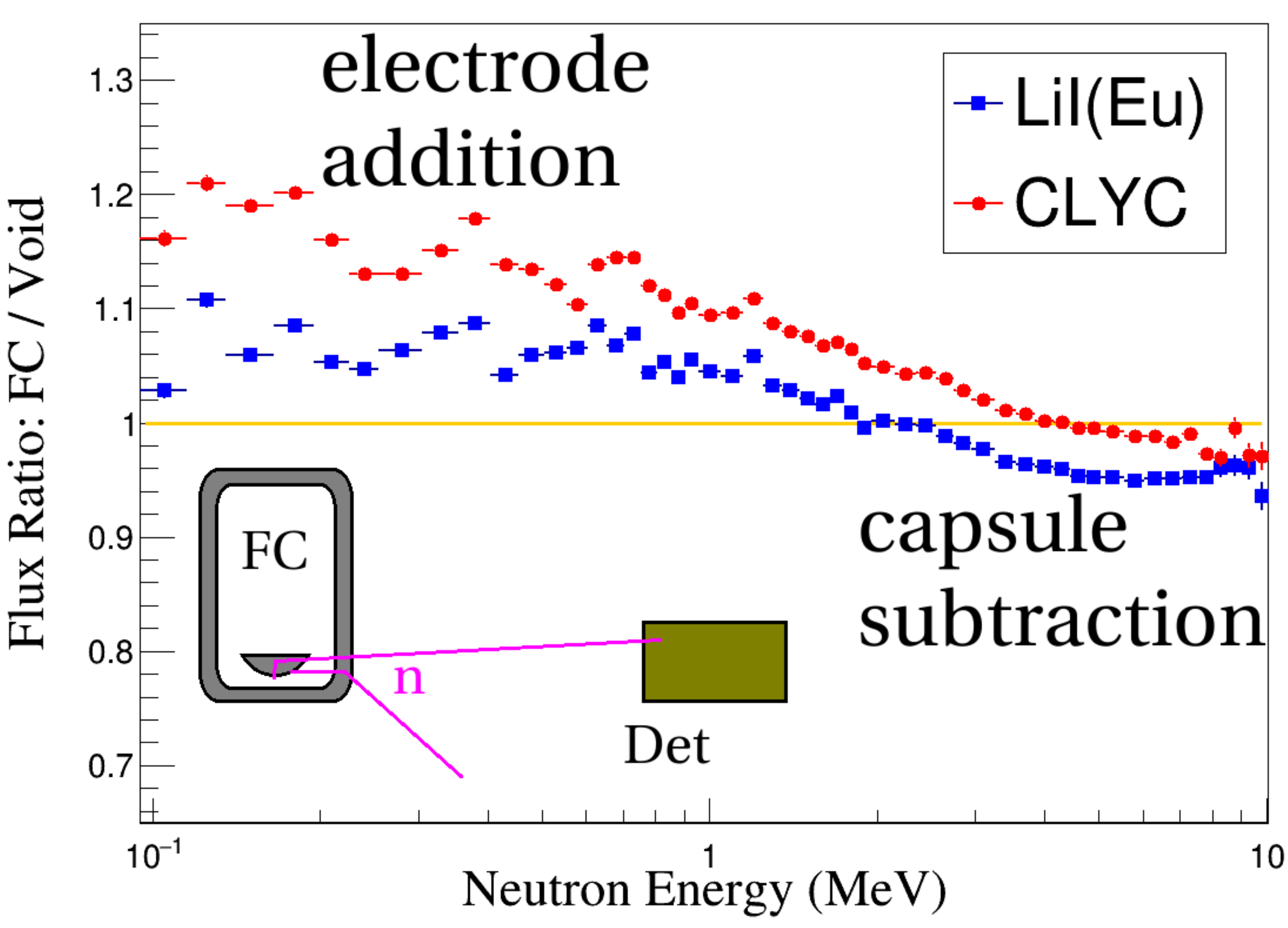}}
\caption{Fission chamber scatter. 
Scattering resonances in steel isotopes cause the fluctuations in flux ratio.
The $^6$LiI(Eu) and C$^6$LYC receive different neutron spectra due to anisotropies in elastic scattering and asymmetric geometry.
}
\label{fig:F1}
\end{figure}

Figure \ref{fig:F1} shows MCNP\textsuperscript{\textregistered} simulation results containing the fission chamber geometry. 
The output, $R_{FC/Void}$, is the energy dependent neutron flux ratio between rooms with and without fission chamber material.
Flux incident upon the detector volume increases with the inclusion of the electrode which acts as a scattering source.
Flux decreases with the inclusion of capsule walls which act as attenuating barriers.
Electrode addition dominates at low energy where large angle scatters are common. 
Capsule subtraction dominates at high energy where elastic scatter is forward-focused.

The correction factor for fission chamber scatter $F_{FC}$ is
\begin{equation}
F_{FC}(E_n) = \frac{1}{R_{FC/Void}(E_n)}.
\end{equation}
The assigned systematic uncertainty for fission chamber scattering effects is 1/4 the maximum correction.
This maximum ensures that systematic uncertainties persist even when electrode addition and capsule subtraction cancel out.


\subsection{\label{ssec:Xtal}Effect of Downscatter}
\begin{figure}
\centerline{\includegraphics[width=0.7\linewidth]{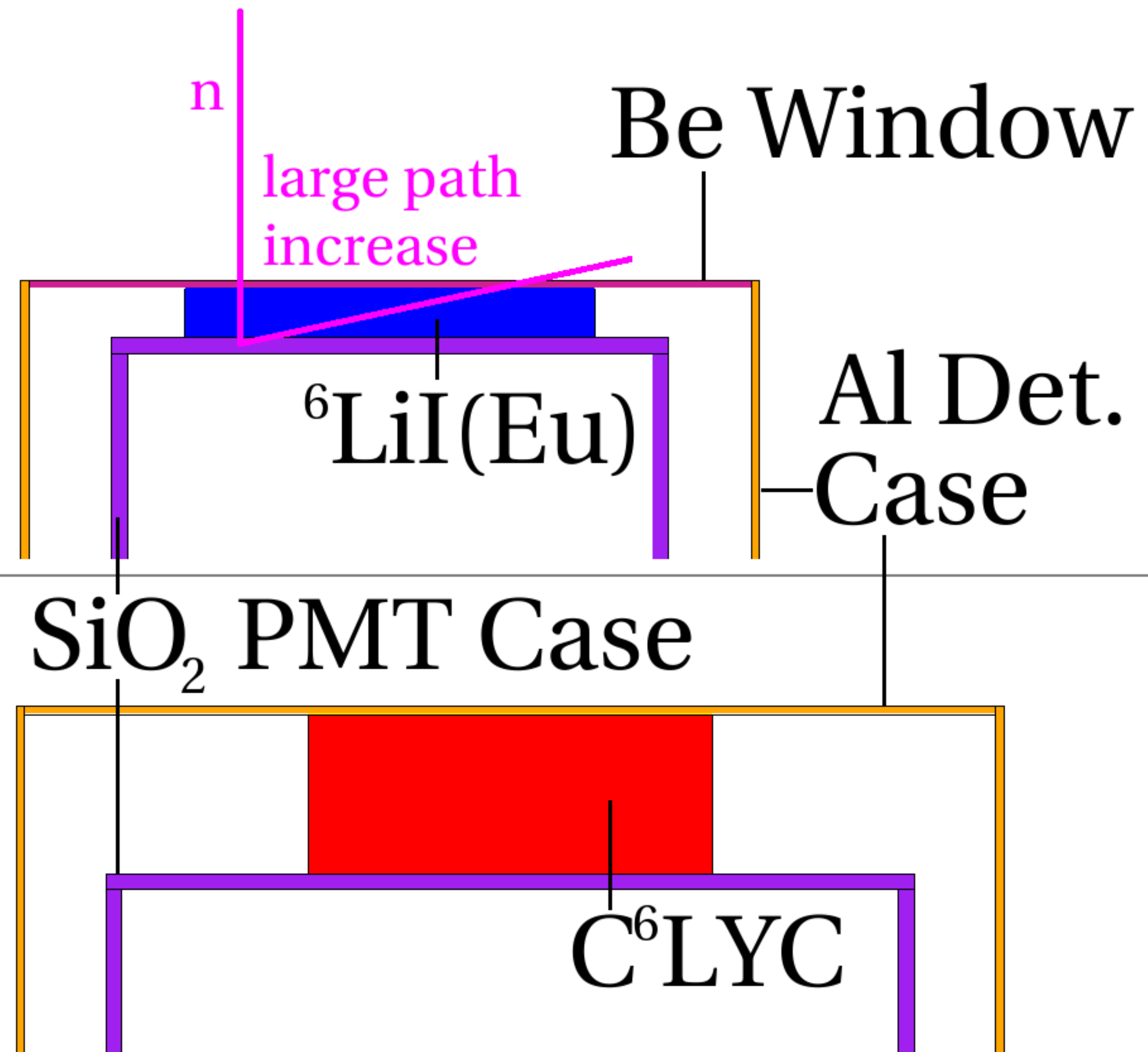}}
\caption{MCNP\textsuperscript{\textregistered} geometry of crystal surroundings.
A neutron backscatters off the PMT glass resulting in a longer path through the $^6$LiI(Eu) crystal.
}
\label{fig:detMCNP}
\end{figure}

Downscatter consists of four effects that modify the neutron flux: path length increase, elastic scatter energy loss, inelastic scatter energy loss, and neutron removal.
These are issues common to all active target experiments where the detector internally hosts the primary reaction alongside an assortment of secondary reactions.
Additionally, the detector casing and PMT usually sit inside the beamline and unavoidably cause neutron scatter.
For example, a neutron can severely increase its path length through the target volume if it scatters in proximity to the crystal as shown in Figure \ref{fig:detMCNP}.
This extended path length increases the target's effective thickness.
The small detector dimensions do not significantly disturb TOF, but downscatter still greatly impacts neutron energy.

\begin{figure}
\centerline{\includegraphics[width=\linewidth]{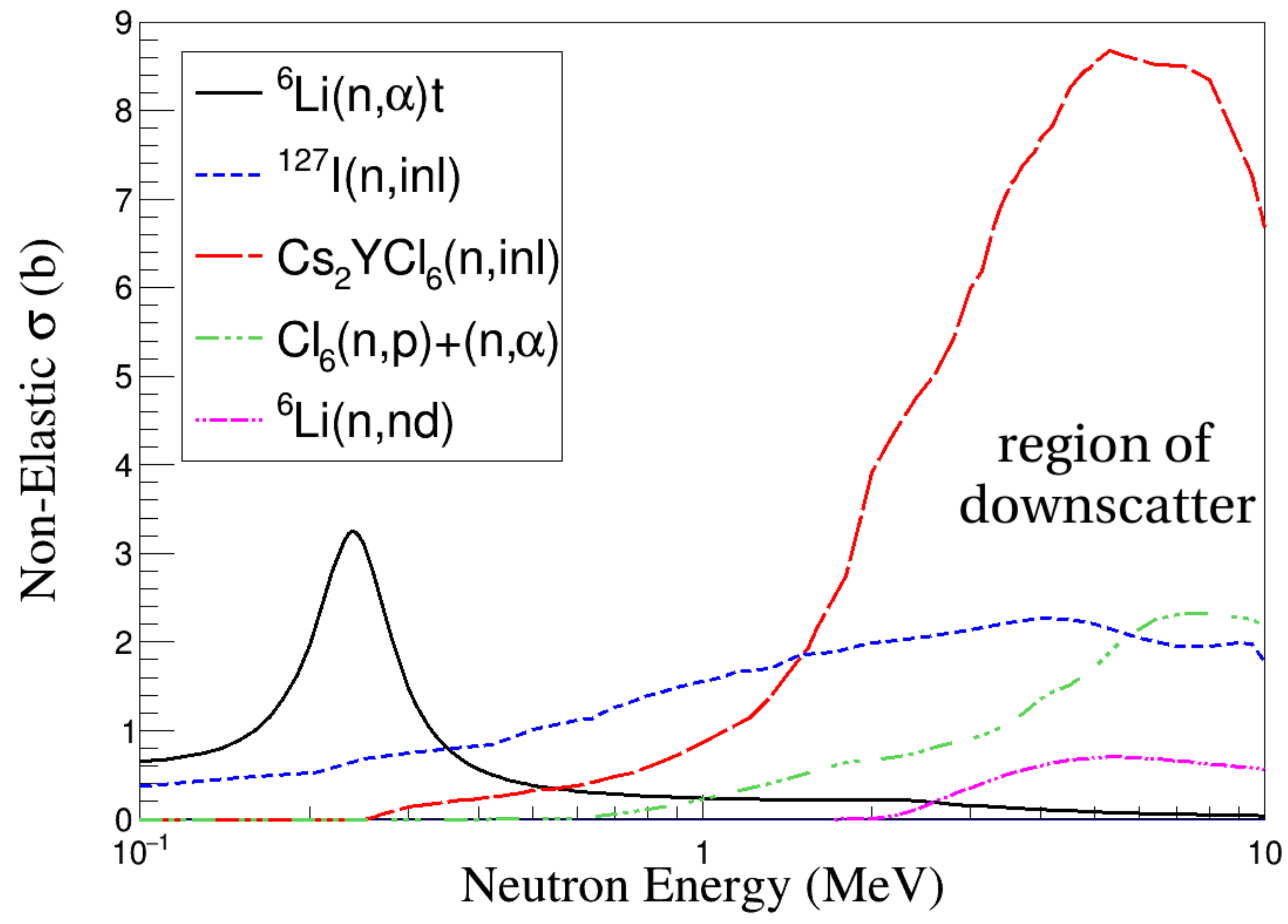}}
\caption{Dominant cross sections in crystal material.
Cross sections for Cs and Cl are weighted by their atomic ratios and isotopic abundances.
Non-pure targets suffer significant neutron attenuation at higher energies.
}
\label{fig:compXS}
\end{figure}
At high incident energies, neutrons lose a lot of their energy via inelastic scattering (n,inl) with $^{127}$I, $^{133}$Cs, $^{89}$Y, and $^{35,37}$Cl.
These other chemicals make scintillation possible, but degrade the quality of an incident neutron beam.
The downscattered neutron beam then continues through the crystal at energies with a larger corresponding $^6$Li(n,$\alpha$)t cross section.
This effect obscures the cross section measurement most at high energies, where as Figure \ref{fig:compXS} shows, the primary $^6$Li(n,$\alpha$)t cross section falls well below that of the secondary reactions available.
%

In C$^6$LYC, the (n,p) and (n,$\alpha$) reactions on $^{35,37}$Cl remove neutrons from the beam, reducing the $^6$Li(n,$\alpha$)t reaction probability altogether. 
The $^{133}$Cs and $^{89}$Y in C$^6$LYC and $^{127}$I in $^6$LiI(Eu) do not have significant (n,p) and (n,$\alpha$) contributions.
Currently there are no experimental measurements of $^{35}$Cl(n,p) from 200 keV to 14 MeV. 
In their work, D'Olympia \textit{et.\ al.\ }show that MCNP\textsuperscript{\textregistered} simulations do not reproduce C$^6$LYC $^{35}$Cl(n,p) spectra well \cite{cite:D׳Olympia2014433}.
As mentioned in Section \ref{sec:expt}, C$^6$LYC has access to the $^{35}$Cl(n,p)$^{35}$S$_{\textrm{g.s.}}$, $^{35}$Cl(n,p)$^{35}$S*, and $^{35}$Cl(n,$\alpha$)$^{32}$P cross sections.
Preliminary results show that $^{35}$Cl(n,p)$^{35}$S$_{\textrm{g.s.}}$ does not match the ENDF/B-VII.1 database \cite{Chadwick20112887} and is the subject of a future article.

\begin{figure}
\centerline{\includegraphics[width=0.75\linewidth]{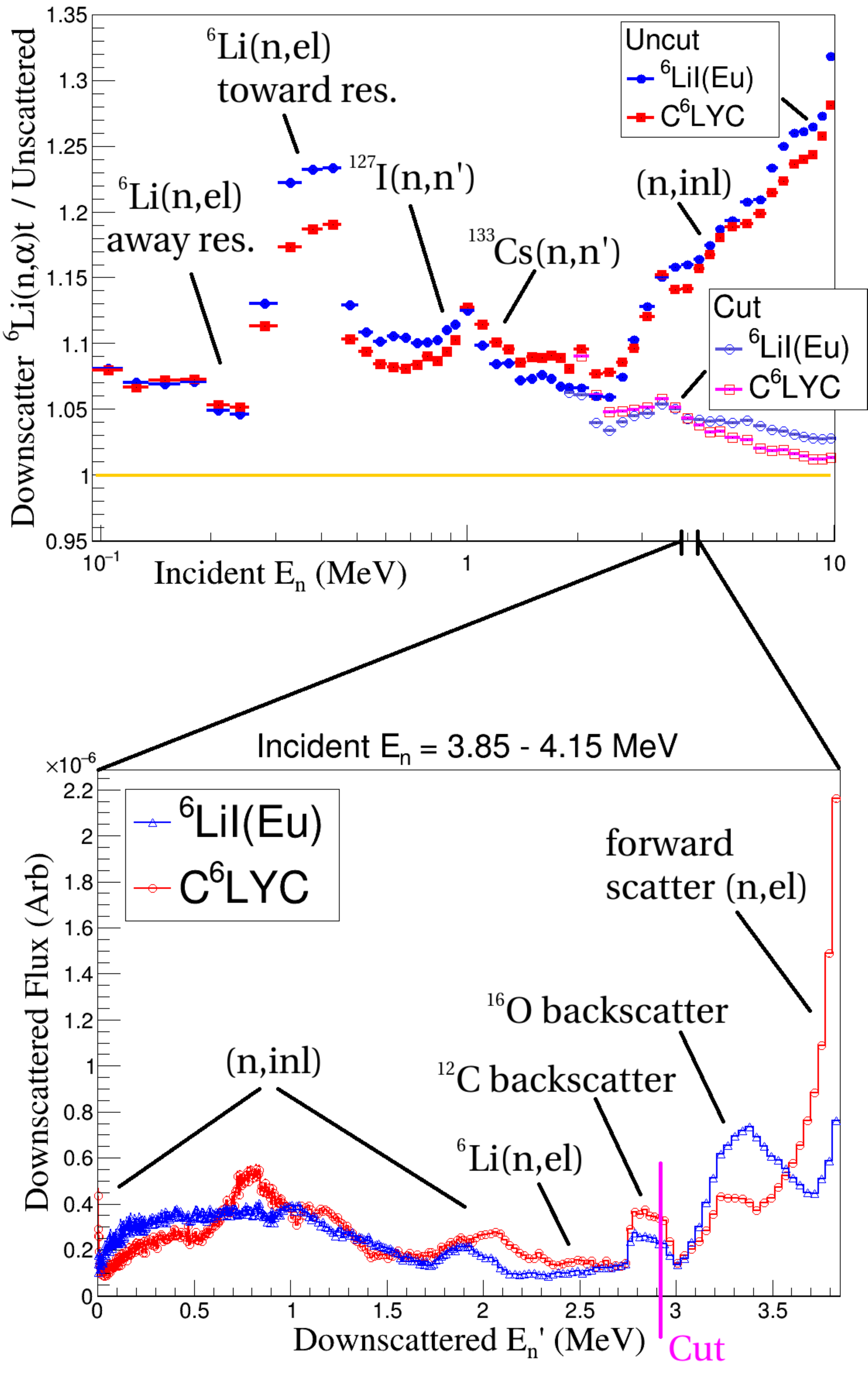}}
\caption{Correction factors for crystal downscatter at all incident energies (top) and downscatter spectrum at 4 MeV (bottom).
The downscatter spectrum is convolved with the $^6$Li(n,$\alpha$)t cross section to generate a correction factor (a single point in the top panel).
For instance, there may not be many downscattered neutrons around 240 keV, but $^6$Li(n,$\alpha$)t is on a resonance at this energy so the correction factor is large. 
However, neutrons that downscatter this low do not induce events with a large enough pulse height to enter the graphical cut of Figure \ref{fig:bksub} (lower bound shown at 2.85 MeV in bottom panel). 
A separate integral, truncated at this low energy bound, produces the correction factor labeled ``Cut'' in the top panel.
}
\label{fig:DownBoth}
\end{figure}

The bottom panel of Figure \ref{fig:DownBoth} shows the result of a typical MCNP\textsuperscript{\textregistered} downscatter simulation. 
Here, nearly monoenergetic neutrons around 4 MeV impinge upon the crystal geometry of Figure \ref{fig:detMCNP} and lose energy in many ways.
This downscatter spectrum is then multiplied by and integrated with the $^6$Li(n,$\alpha$)t cross section to obtain a correction factor at each energy as shown in the top panel.
The explicit expression for this correction factor is
\begin{equation}
F_{down}(E_{n}) = \frac{ \int dE_n' \sigma(E_n') n_{down}(E_n') } 
  {\sigma(E_n) n(E_n)},
\label{eq:P_down}
\end{equation}
where $n_{down}$ is the downscattered flux of the bottom panel of Figure \ref{fig:DownBoth} and $n(E_n)$ is the incident unscattered flux.
There are a lot of additional downscatter $^6$Li(n,$\alpha$)t reactions at high energies, but (n,inl) often puts the neutron energy outside the graphical cut of Figure \ref{fig:bksub}.
Events outside the cut do not contribute to the cross section correction.

The uncertainty assigned to the combination of simulation geometry and cross sections is large: 1/3 the correction. This includes uncertainty associated with the subtraction of incident flux and addition of epithermal flux.

\subsection{\label{ssec:resp}High Energy Particle Leakage }
In their work, Murray and Schmitt \cite{J.PR.115.1707.59} suggest a correction to the data to account for the loss of alphas and tritons from the faces of thin crystals.
These effects are most prominent at high incident neutron energies where the recoiling ejectiles travel a large distance.

Particle leakage is not a straightforward effect to estimate.
Cross sections depend on outgoing particle angles which govern the two-body kinematics.
Even if one or both ions leave the detector, they may deposit enough energy prior to leaving such that the event pulse height enters the graphical cut of Figure \ref{fig:bksub}.
Photoluminescence determines pulse height, but it is a complex condensed matter process and the corresponding light response curves require prior measurement.
At least for $^6$LiI(Eu), there are data to show these curves differ vastly between electrons, alphas, protons, and other various charged particles \cite{J.NP_A.330.1.197910}.
These charged particles will have different recoil distances for equivalent initial kinetic energies.
There is some theoretical background on ion ranges in matter, but they too require experimental verification. 
Fortunately, many tables exist that provide readily accessible projected-range data for almost all target and incident ions \cite{Ziegler20101818}.
Finally, experimental timing and pulse height resolutions obscure the boundaries between escaping and stopping ions.
A newly created detector response simulation code called \texttt{detResp.C} addresses each of these phenomena and approximates event loss due to particle leakage specifically for the $^6$LiI(Eu) and C$^6$LYC crystals.

\begin{figure}
\includegraphics[width=\linewidth]{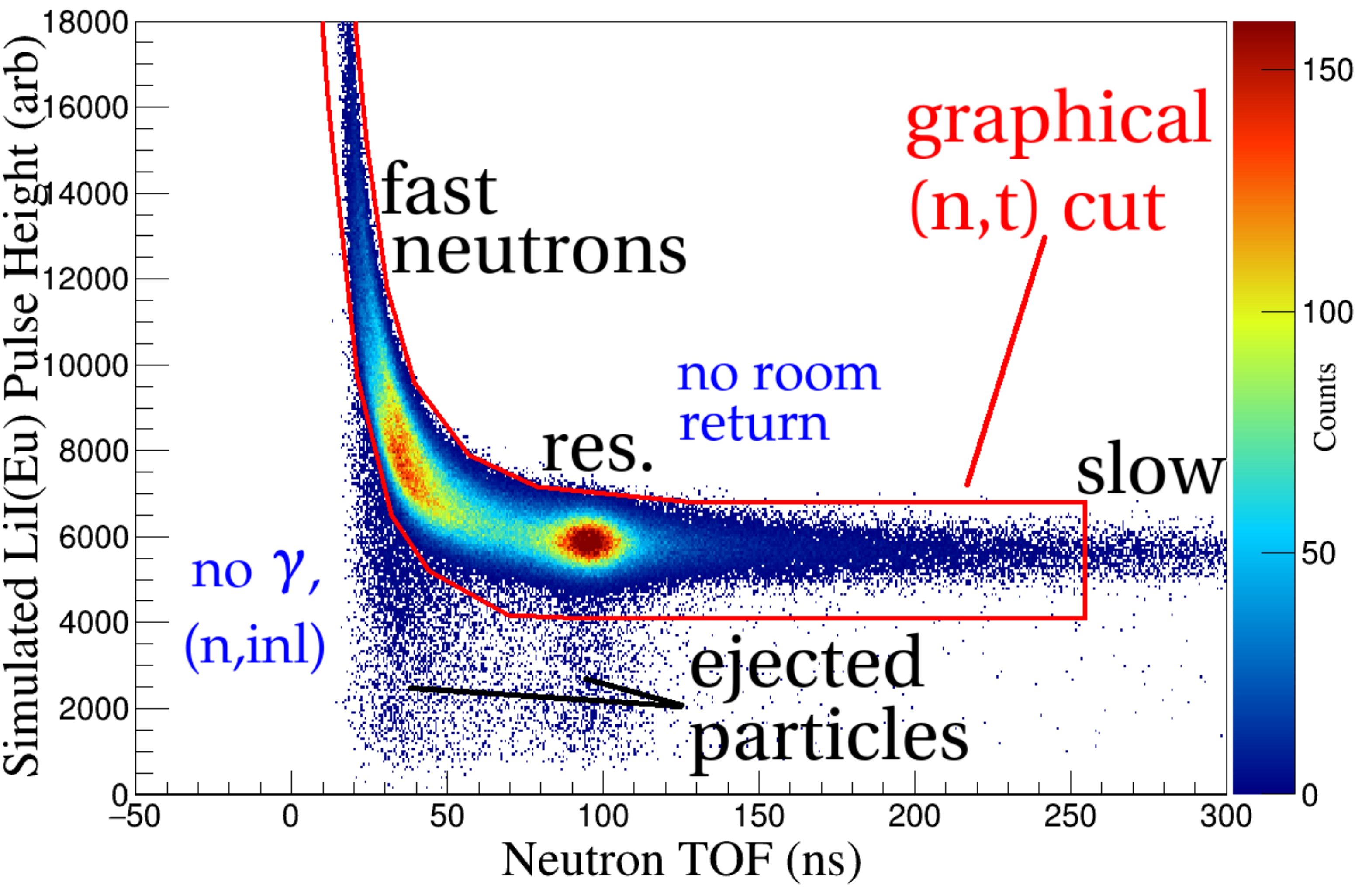}
\caption{Simulation of $^6$LiI(Eu) response to $^{252}$Cf fission neutron spectrum.
This \texttt{detResp.C} simulation omits the $\gamma$-ray flash, room scattering geometry, and other inelastic reactions besides $^6$Li(n,$\alpha$)t, but otherwise has all the features of the background subtracted spectrum of Figure \ref{fig:bksub}.}
\label{fig:simLiI}
\end{figure}
Figure \ref{fig:simLiI} shows \texttt{detResp.C} simulated pulse height vs.\ TOF spectrum for a $^{252}$Cf neutron spectrum incident upon a $^6$LiI(Eu) detector.
Equation \eqref{eq:En} converts random samples of neutron energy from the Mannhart spectrum \cite{cite:cfSpec} into TOF.
Reaction locations inside the cylindrical crystal geometry are uniformly distributed, neglecting self-shielding.
Random samples of $^6$Li(n,$\alpha$)t angle differential cross section \cite{Chadwick20112887} govern triton ejection angles and correlated alpha ejection angles.
Elementary recoil kinematics \cite{cite:krane} determine initial ejectile recoil energies from emission angles.
Stopping curves \cite{Ziegler20101818} determine energy deposition of each particle before, or if, it escapes.
Renormalized detector response curves \cite{J.NP_A.330.1.197910} determine pulse height from particle energy deposition.
Experimental timing and pulse height resolutions simulate fluctuations so that any given event has a finite probability to escape the graphical cut. 
Analysis of fabricated data incorporates the same graphical cut as Figure \ref{fig:bksub} to determine event escape probability based on simulated TOF and pulse height. 

\begin{figure}
\centerline{\includegraphics[width=\linewidth]{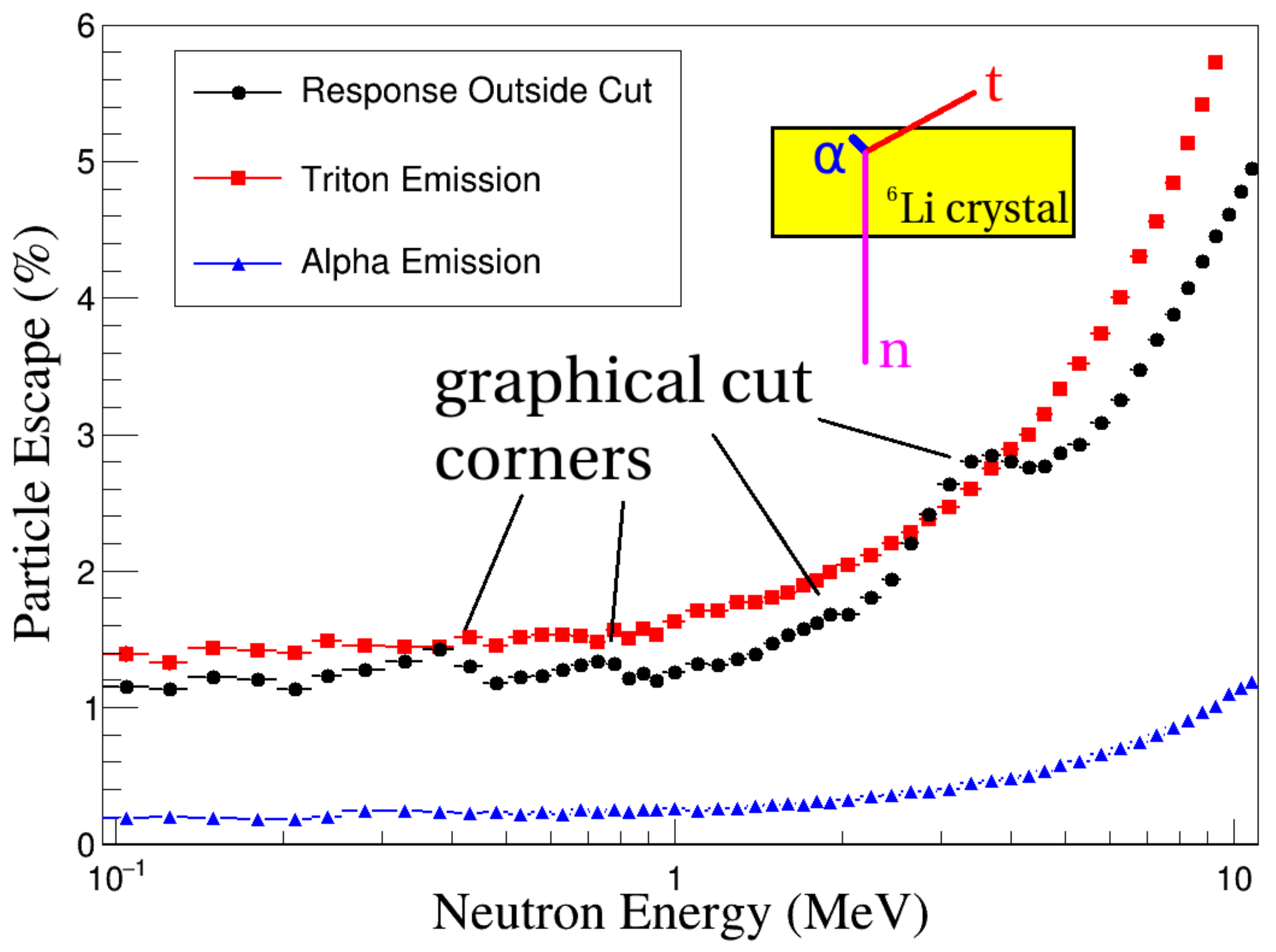}}
\caption{Triton, alpha, and response escape in $^6$LiI(Eu). 
Sharp corners in the graphical cut impose noticeable shoulder features in the response cut escape. 
No such feature exists for the triton and alpha escape curves since escape does not rely on the graphical cut but rather on detector geometry.
}
\label{fig:esc}
\end{figure}

Using results of the previous simulation, Figure \ref{fig:esc} plots escape probabilities vs.\ neutron energy.
Triton leakage is more likely than alpha leakage because ion range is inversely proportional to $Z_i^2 m_i$, where $Z_i$ is ion charge, and $m_i$ is ion mass.
Therefore tritons have roughly 5.3 times the range as alpha particles.
Murray and Schmitt \cite{J.PR.115.1707.59} propose a 6\% correction at neutron energy 8 MeV for a 2.5 mm thick $^6$LiI(Eu) crystal and express that this leakage effect increases with incident neutron energy.
The simulation of a slightly thicker $^6$LiI(Eu) crystal has a comparable 5.5\% combined alpha and triton leakage at 8 MeV.
However, not all ion-escape events have total pulse height outside the graphical cut due to its finite acceptance width. 

The range curves of alphas, tritons, and protons in C$^6$LYC are very similar to those of $^6$LiI(Eu).
The event loss for the C$^6$LYC crystal is not shown in Figure \ref{fig:esc}, but is approximately a factor of 10/3 smaller since the crystal is a factor of 10/3 times thicker than the $^6$LiI(Eu).

The correction factor for particle escape $F_{Esc}$ is
\begin{equation}
F_{Esc}(E_{n}) = \frac{1} {1 - P_{Esc}(E_{n}) } ,
\end{equation}
where $P_{Esc}$ is the fraction of events that have pulse heights outside the graphical cut at incident neutron energy $E_n$, shown in Figure \ref{fig:esc}.
The assigned systematic uncertainty is 1/4 the correction.


\section{\label{sec:xs}Corrected Cross Section and Discussion}

The correction factors for fission chamber scatter $F_{FC}$, crystal downscatter $F_{down}$, and particle escape $F_{esc}$ modify Equation \eqref{eq:xs} as follows:
\begin{equation}
\sigma_{corr}(E_n) = \frac{n_n(E_n) F_{FC} F_{down} F_{esc}}{n_f(E_n) N_{f,tot} \phi_{det} N_{^6\textrm{Li}} }.
\label{eq:xsCorr}
\end{equation}
Section \ref{ssec:Room} shows that this corrected cross section, $\sigma_{corr}$, is valid for $E_n>$ 245 keV since room return is absent.

\begin{figure}
\centerline{\includegraphics[width=\linewidth]{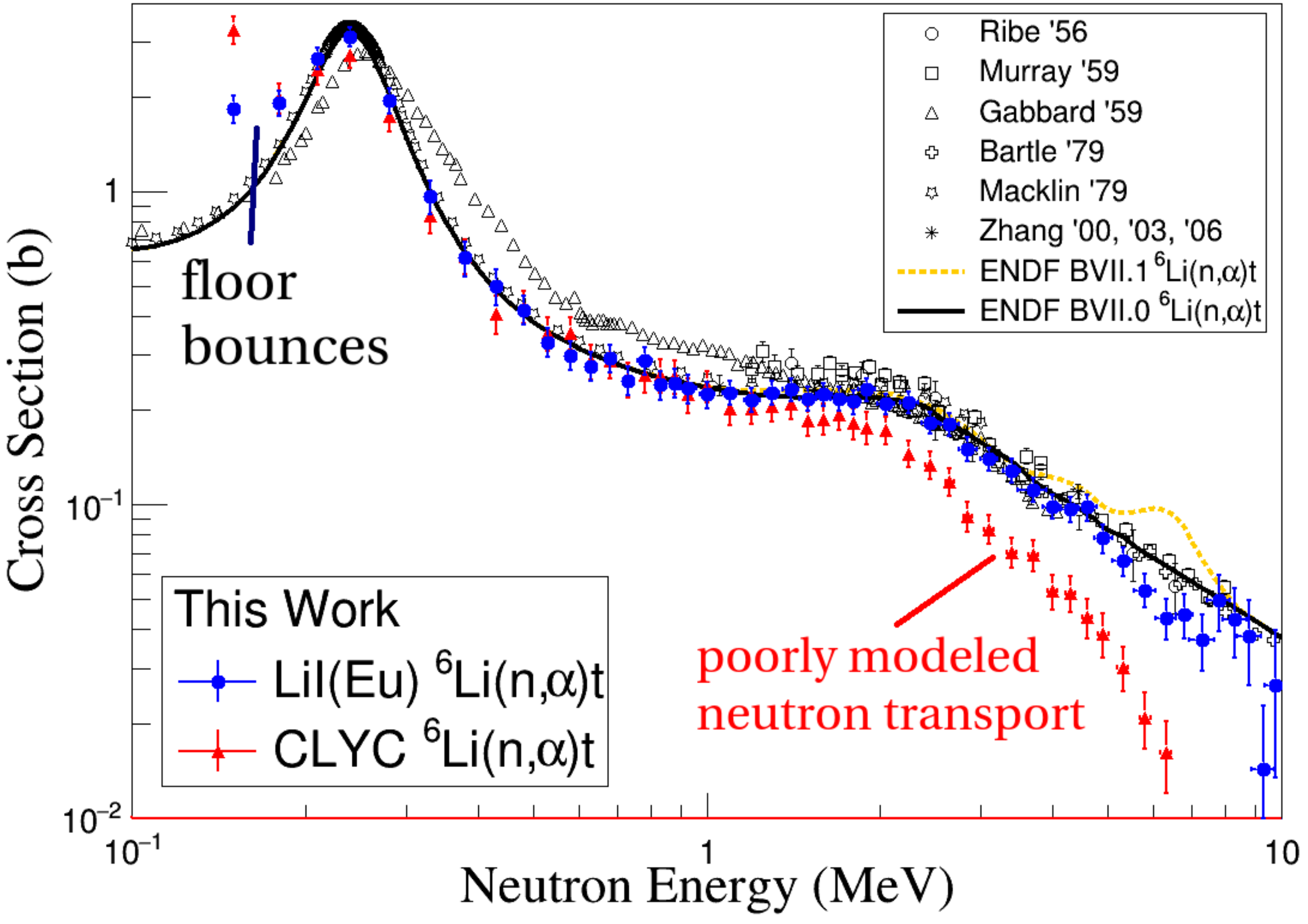}}
\caption{Experimental $^6$Li(n,$\alpha$)t absolute cross section with correction factors included.
The data shown was collected exclusively with detectors in the 125 cm far configuration.
The upturn below the 240 keV resonance is the result of room return, specifically single scatters off the floor. 
Inadequate $^{35,37}$Cl(n,p) and (n,inl) database values result in poorly modeled neutron transport simulations for C$^6$LYC.
}
\label{fig:finalXS}
\end{figure}

\begin{figure}
\centerline{\includegraphics[width=\linewidth]{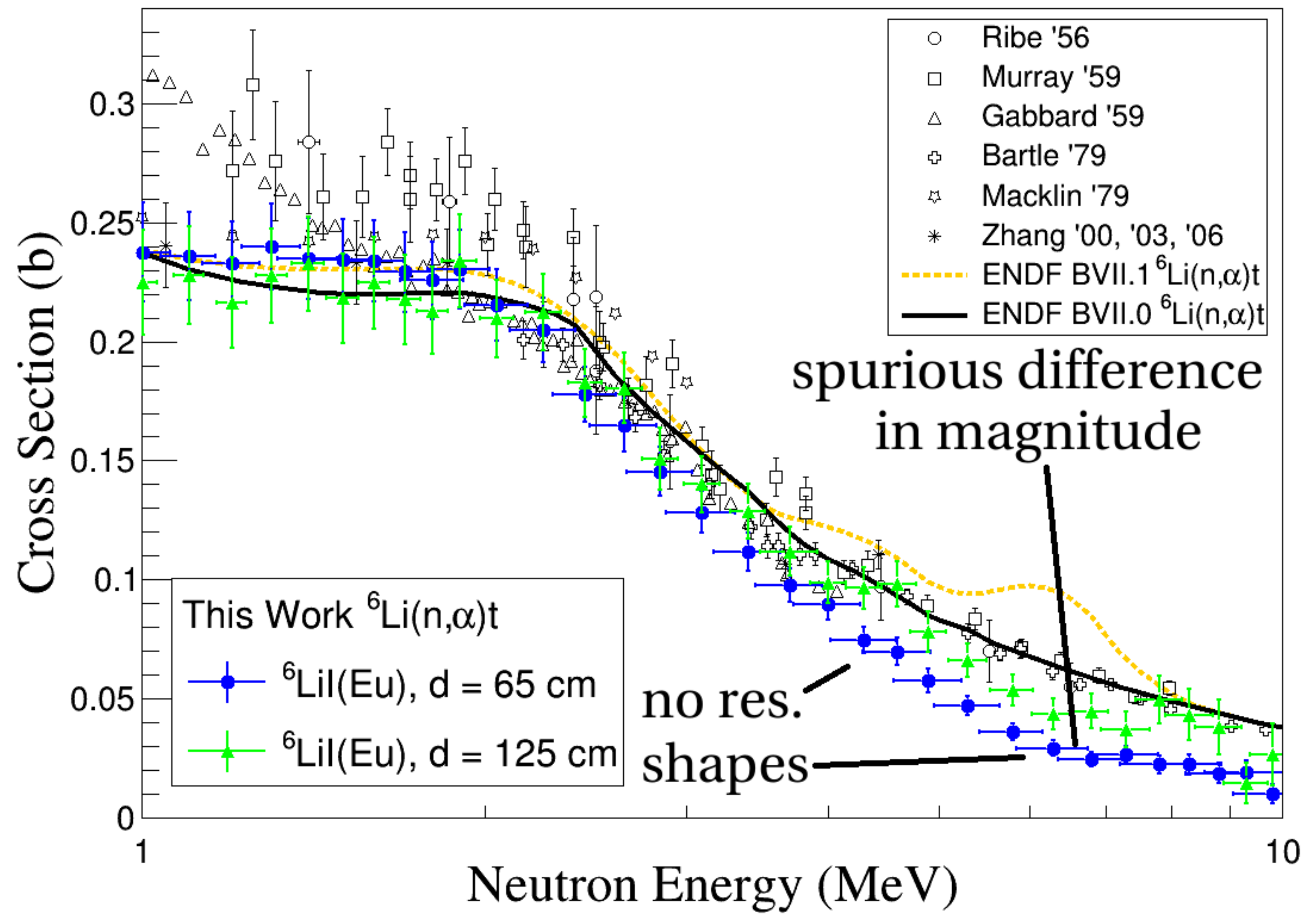}}
\caption{$^6$Li(n,$\alpha$)t cross section measured with $^6$LiI(Eu) detector in the near and far configurations.
An unresolved factor causes a discrepancy in cross section magnitude at high energies.
However, the large number of counts in the near configuration data reduces statistical fluctuations and reveals that there are no outstanding resonances at 4.2 and 6.5 MeV.
}
\label{fig:closeXS}
\end{figure}

Figure \ref{fig:finalXS} shows the corrected $^6$Li(n,$\alpha$)t cross section.
Unlike previous experimental and simulated plots, Figures \ref{fig:finalXS}, \ref{fig:closeXS}, and Table \ref{tab:xs} display results for the detectors in the 125 cm far configuration. 
The $^6$LiI(Eu) data match previous experiments and the ENDF/B-VII.0 evaluation \cite{Chadwick20062931} from 245 keV to 10 MeV reasonably well in both magnitude and shape.
However, the C$^6$LYC data had issues above 1.5 MeV likely due to poorly modeled simulation of neutron transport inside the crystal since there are few experimental cross section measurements for $^{133}$Cs, $^{89}$Y, and $^{35,37}$Cl.
In certain cases, the MCNP\textsuperscript{\textregistered} transport simulation relies exclusively on prior calculations by the reaction codes EMPIRE \cite{HERMAN20072655} and SAMMY \cite{SAMMY}.
ENDF evaluations utilizing these codes do include emitted particle angular distributions, but depend on phenomenological fits and parameterizations of many models and few experiments.
Since $^6$Li has only a 9.5\% atom fraction in C$^6$LYC and the calculated cross sections do not have much experimental support, it is not surprising that quantitative agreement for the $^6$Li(n,$\alpha$)t measurement is poor.
The chlorine isotopes are especially sensitive, since they have the highest isotopic abundance present but the least experimental data.
For instance, a poorly simulated magnitude of $^{35}$Cl(n,p) and (n,inl) could underestimate neutron flux attenuation; which in reality brings down the $^6$Li(n,$\alpha$)t cross section measurement. 


At high incident neutron energies the multitude of systematic uncertainties become difficult to manage. 
Near relativistic speeds the energy determination is increasingly sensitive to the measurement of distance from fission chamber to detector. 
Both neutron flux and $^6$Li(n,$\alpha$)t cross section are low at high energy. 
When a high energy $^6$Li(n,$\alpha$)t event does occur, the particle ejection probability is high and the shape of the graphical cut of Figure \ref{fig:bksub}  may not optimally select $^6$Li(n,$\alpha$)t events while excluding downscattered events and other reactions. 
Figure \ref{fig:DownBoth} shows that this cut makes the difference between a 4\% and 15\% downscatter correction at 4 MeV.
Finally, the steel fission chamber capsule consists of many isotopes for which there are not many cross section measurements at high energies. 
The scarce data, along with machining tolerances make the fission chamber scatter difficult to simulate.

The ENDF/B-VII.1 evaluation adds two large resonances to the $^6$Li(n,$\alpha$)t cross section at 4.2 MeV and 6.5 MeV.
The measurements shown in Figure \ref{fig:finalXS} do not reveal these resonances, but the statistical fluctuations at these energies might be large enough to conceal the structure.
The fluctuations are large in Figure \ref{fig:finalXS} because the data shown are from the iteration where the detectors were in the far configuration where solid angle is small and the number of $^6$Li(n,$\alpha$)t counts are low.
To further investigate the existence of these resonances, Figure \ref{fig:closeXS} shows cross section results for $^6$LiI(Eu) in the near configuration where the solid angle and number of counts are larger.
At high energies, the cross section measurement of the near configuration has a discrepancy in magnitude which may be due to uncertainties in path length or the shape of the graphical cut.
However, any large resonances would still appear with perhaps a slight shift in centroid energy. 
Therefore, the smaller statistical fluctuations of this iteration resolve small scale shape and confirm that there are no large resonances at 4.2 MeV and 6.5 MeV. 

Table \ref{tab:xs} reports the $^6$Li(n,$\alpha$)t cross section measured by $^6$LiI(Eu) in the far configuration which has the most reliable magnitude and overall shape.
\begin{table}
\centering
\caption{Final Corrected $^6$Li(n,$\alpha$)t Cross Section from $^6$LiI(Eu) in the Far Configuration}
\resizebox{\linewidth}{!}{%
\begin{tabular}{c c||c c}
  \hline\hline
  E (MeV) & $\sigma$ (b) & E (MeV) & $\sigma$ (b)\\
  \hline
0.24   $\pm$0.0047 &3.12  $\pm$0.23  &1.9  $\pm$0.064 &0.234  $\pm$0.020  \\
0.28   $\pm$0.0058 &1.95  $\pm$0.18  &2.05 $\pm$0.068 &0.210  $\pm$0.016  \\
0.33   $\pm$0.0070 &0.961 $\pm$0.12  &2.25 $\pm$0.081 &0.212  $\pm$0.016  \\
0.38   $\pm$0.0086 &0.613 $\pm$0.077 &2.45 $\pm$0.084 &0.183  $\pm$0.014  \\
0.43   $\pm$0.010  &0.498 $\pm$0.064 &2.65 $\pm$0.094 &0.181  $\pm$0.015  \\
0.48   $\pm$0.012  &0.419 $\pm$0.044 &2.85 $\pm$0.11  &0.151  $\pm$0.013  \\
0.53   $\pm$0.013  &0.330 $\pm$0.035 &3.1  $\pm$0.11  &0.140  $\pm$0.012  \\
0.58   $\pm$0.015  &0.299 $\pm$0.032 &3.4  $\pm$0.12  &0.129  $\pm$0.011  \\
0.63   $\pm$0.016  &0.277 $\pm$0.029 &3.7  $\pm$0.21  &0.112  $\pm$0.010  \\
0.68   $\pm$0.018  &0.293 $\pm$0.030 &4.0  $\pm$0.15  &0.0988 $\pm$0.0088 \\
0.73   $\pm$0.019  &0.249 $\pm$0.026 &4.3  $\pm$0.16  &0.0964 $\pm$0.0089 \\
0.78   $\pm$0.023  &0.289 $\pm$0.029 &4.6  $\pm$0.18  &0.0981 $\pm$0.0094 \\
0.83   $\pm$0.023  &0.241 $\pm$0.025 &4.9  $\pm$0.18  &0.0781 $\pm$0.0084 \\
0.88   $\pm$0.025  &0.245 $\pm$0.026 &5.3  $\pm$0.21  &0.0662 $\pm$0.0069 \\
0.93   $\pm$0.029  &0.235 $\pm$0.025 &5.8  $\pm$0.25  &0.0533 $\pm$0.0066 \\
1.0025 $\pm$0.030  &0.225 $\pm$0.022 &6.3  $\pm$0.27  &0.0434 $\pm$0.0064 \\
1.1    $\pm$0.033  &0.228 $\pm$0.021 &6.8  $\pm$0.23  &0.0445 $\pm$0.0073 \\
1.2    $\pm$0.037  &0.216 $\pm$0.019 &7.3  $\pm$0.37  &0.0369 $\pm$0.0075 \\
1.3    $\pm$0.040  &0.228 $\pm$0.020 &7.8  $\pm$0.36  &0.0495 $\pm$0.010  \\
1.4    $\pm$0.046  &0.233 $\pm$0.020 &8.3  $\pm$0.41  &0.0431 $\pm$0.011  \\
1.5    $\pm$0.049  &0.218 $\pm$0.019 &8.8  $\pm$0.42  &0.0380 $\pm$0.012  \\
1.6    $\pm$0.050  &0.225 $\pm$0.019 &9.3  $\pm$0.41  &0.0143 $\pm$0.0085 \\
1.7    $\pm$0.057  &0.218 $\pm$0.019 &9.8  $\pm$0.42  &0.0265 $\pm$0.013  \\
1.8    $\pm$0.057  &0.213 $\pm$0.018 & & \\
  \hline\hline
\end{tabular}}
\label{tab:xs}
\end{table}

\section{\label{sec:conc}Conclusion }
We have presented a new measurement of the $^6$Li(n,$\alpha$)t cross section in the neutron energy range of 245 keV to 10 MeV, which to our knowledge is the first of such measurements utilizing a fission chamber.
The measurements with $^6$LiI(Eu) disagree with the addition of two resonances at 4.2 and 6.5 MeV by the ENDF/B-VII.1 evaluation. 
Correction factors for neutron downscatter in both the fission chamber and $^6$LiI(Eu) crystal as well as particle leakage seem tractable within this energy range. 
However, the correction factor for neutron downscattering in the C$^6$LYC crystal does not bring the $^6$Li(n,$\alpha$)t measurement into quantitative agreement with previous results. 
This inconsistency is likely the result of an unmeasured $^{35}$Cl or $^{37}$Cl cross section which supports previous simulation discrepancies of other authors.

\section{\label{sec:ackn}Acknowledgements }
This work was performed with the support of the DOE NNSA Stewardship Science Graduate Fellowship under cooperative agreement number DE-NA0002135 in joint affiliation with Los Alamos National Laboratory.
It is a pleasure to thank to L.\ A.\ Bernstein, T.\ Brown, M.\ Simanovskaia, and A.\ Lewis for stimulating conversations and suggestions. 

\bibliography{pub}

\end{document}